\documentclass[sigconf]{acmart}

\usepackage{enumitem}
\usepackage{color}
\usepackage{float}
\usepackage{stfloats}
\usepackage{soul}
\usepackage{tabularx}
\usepackage{multirow}
\usepackage{booktabs}
\usepackage{hyperref}
\usepackage{xcolor}
\usepackage{makecell}

\AtBeginDocument{%
  \providecommand\BibTeX{{%
    \normalfont B\kern-0.5em{\scshape i\kern-0.25em b}\kern-0.8em\TeX}}}

\copyrightyear{2023}
\acmYear{2023}
\setcopyright{rightsretained}
\acmConference[UIST '23]{The 36th Annual ACM Symposium on User Interface Software and Technology}{October 29-November 1, 2023}{San Francisco, CA, USA}
\acmBooktitle{The 36th Annual ACM Symposium on User Interface Software and Technology (UIST '23), October 29-November 1, 2023, San Francisco, CA, USA}
\acmDOI{10.1145/3586183.3606770}
\acmISBN{979-8-4007-0132-0/23/10}

\newcommand{\sysname}{Papeo}

\begin{document}

%%
%% The "title" command has an optional parameter,
%% allowing the author to define a "short title" to be used in page headers.
%\title{Papeos: Augmenting Research Papers with Talk Videos}\titlenote{\colorbox{yellow}{\textcolor{black}{\href{https://papeo.app/demo}{Click here to open the Papeo version of this document! \url{https://papeo.app/demo}}}}}
\title{Papeos: Augmenting Research Papers with Talk Videos}\titlenote{\href{https://papeo.app/demo}{Click here to open the Papeo version of this document: \textcolor{blue}{\url{https://papeo.app/demo}}}}

%%
%% The "author" command and its associated commands are used to define
%% the authors and their affiliations.
%% Of note is the shared affiliation of the first two authors, and the
%% "authornote" and "authornotemark" commands
%% used to denote shared contribution to the research.
\author{Tae Soo Kim}
\authornote{Work completed during a researcher internship at Semantic Scholar Research, Allen Institute for AI.}
\affiliation{%
  \institution{School of Computing, KAIST}
  \city{Daejeon}
  \country{Republic of Korea}}
\email{taesoo.kim@kaist.ac.kr}

\author{Matt Latzke}
\affiliation{%
  \institution{Allen Institute for AI}
  \streetaddress{2157 N Northlake Way \#110}
  \city{Seattle}
  \state{WA}
  \postcode{98103}
  \country{USA}
}
\email{mattl@allenai.org}

\author{Jonathan Bragg}
\affiliation{%
  \institution{Allen Institute for AI}
  \city{Seattle}
  \state{WA}
  \country{USA}
}
\email{jbragg@allenai.org}

\author{Amy X. Zhang}
\affiliation{%
  \institution{University of Washington}
  \city{Seattle}
  \state{WA}
  \country{USA}
}
\email{axz@cs.uw.edu}

\author{Joseph Chee Chang}
\affiliation{%
  \institution{Allen Institute for AI}
  \city{Seattle}
  \state{WA}
  \country{USA}
}
\email{josephc@allenai.org}

%%
%% By default, the full list of authors will be used in the page
%% headers. Often, this list is too long, and will overlap
%% other information printed in the page headers. This command allows
%% the author to define a more concise list
%% of authors' names for this purpose.
\renewcommand{\shortauthors}{Kim et al.}

%%
%% The abstract is a short summary of the work to be presented in the
%% article.
\begin{abstract}

Research consumption has been traditionally limited to the reading of academic papers---a static, dense, and formally written format. Alternatively, pre-recorded conference presentation videos, which are more dynamic, concise, and colloquial, have recently become more widely available but potentially under-utilized. In this work, we explore the design space and benefits for combining academic papers and talk videos to leverage their complementary nature to provide a rich and fluid research consumption experience. Based on formative and co-design studies, we present \textbf{Papeos}, a novel reading and authoring interface that allow authors to augment their \textbf{pap}ers by segmenting and localizing talk vid\textbf{eos} alongside relevant paper passages with automatically generated suggestions. With Papeos, readers can visually skim a paper through clip thumbnails, and fluidly switch between consuming dense text in the paper or visual summaries in the video. In a comparative lab study (n=16), Papeos reduced mental load, scaffolded navigation, and facilitated more comprehensive reading of papers.

\end{abstract}

%%
%% The code below is generated by the tool at http://dl.acm.org/ccs.cfm.
%% Please copy and paste the code instead of the example below.
%%



\maketitle

\section{Introduction}

Research progress is driven by the consumption of prior research. 
This is most commonly performed through reading published academic papers that are written in a rigid, dense, and formal fashion to ensure clarity, reproducibility, and to fit within a tight page limit constraint. 
Alternatively, another common way researchers learn about prior research is through conference presentations.
Presentations or talks are typically more concise and colloquial, and can contain rich and dynamic content that cannot be included in the paper, such as screencasts of user interfaces, animated figures, and progressive diagrams.

%% Teaser
\begin{figure}
  \centering
  \includegraphics[width=1.00\columnwidth]{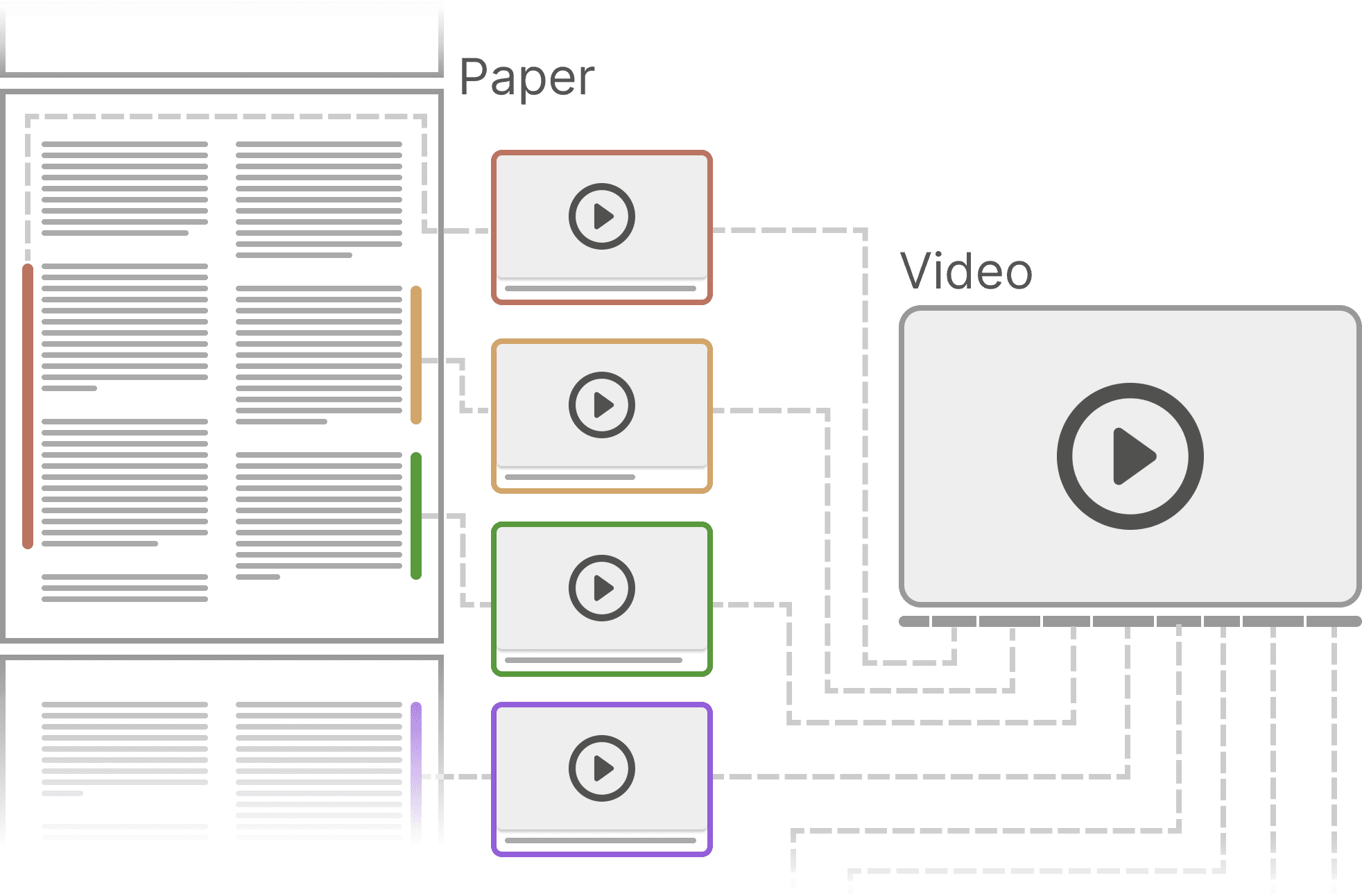}
  \caption{\sysname{}s augment academic papers by linking relevant passages and segments of authors' talk videos. Video segments are presented as margin notes that are localized and color-coded next to relevant passages. In \sysname{}s, users can fluidly switch between consuming the dense and detailed paper text, and the typically more concise and easier to understand talk video---providing a new scholarly reading experience. See system screenshots in Figures~\ref{fig:reader} and \ref{fig:reader-features}.}
  \label{fig:teaser}
\end{figure}

Traditionally, author presentations were typically only available to attendees and only during conferences. 
However, for archival purposes~\cite{amir2001towards} or to accommodate remote participants (e.g., during the COVID-19 pandemic, to be more inclusive~\cite{chihybrid}, and to reduce carbon emissions~\cite{higham2020decarbonising}), recordings and pre-recordings of conference talks (i.e., talk videos) have become more widely available across different fields in recent years.
Prior work, such as in psychology and education, has found various benefits in a personal and multimedia communication style (i.e., videos and dialogues) over formal and technical text, including positive social context and experiences \cite{kang2012effects}, lowered cognitive load and increased interest \cite{Mayer2004APE}, and improved comprehension when multiple alternative explanations were available \cite{Ainsworth2006DeFTAC}. However, prior HCI research has also showed that carefully designed interfaces are crucial for users to consume multiple formats without being overwhelmed~\cite{grossman2015your}. In this work, we build on prior theoretical and HCI research to explore the design space for combining research papers and talk videos into a cohesive reading experience by investigating the perspectives of both paper authors and readers.

Talk videos differ from papers in format and content, and this can serve to address various challenges in research consumption.
Specifically, while reading papers allows scholars to dig deep into all the details of a prior work, the process can be cognitively demanding as scholars must disentangle meaning from complex written explanations~\cite{bazerman1985physicists}.
This process is further complicated as researchers may lack the background knowledge required to understand the explanations or due to variability in the quality of the writing~\cite{ozuru2009text, otero2014psychology}.
Even further, to keep pace with the rapidly expanding literature, researchers are increasingly pressured to skim papers, and they attempt to gain a high-level understanding from scattered fragments of writing~\cite{maxwell1972skimming, hillesund2010digital}.
In contrast, a talk video may present visuals that can help illustrate complex explanations~\cite{chun1996facilitating, fu2022doc2ppt, rowley2004different} and, due to their wider audience, focus less on specialized concepts or background knowledge while using simpler language~\cite{rowley2005rhetoric, dang2022corpus}.
Furthermore, as talk videos typically do not contain all the details, they can present scholars with a concise and easy-to-understand overview of the corresponding papers~\cite{lev2019talksumm, carter2003analysing}.

Despite the various ways in which talk videos can complement paper reading, these two formats remain largely disconnected. Readers have to choose between using either the talk video or the paper as their primary way to consume prior work, and cognitive costs to switch between the two formats could be prohibitively high. For example, if a scholar watching a talk video wants to find a specific implementation detail for a machine learning model that was omitted in the video, they must search through pages in the paper to find the corresponding passage.
Similarly, when reading a paper about an interactive user interface, it can also be costly for a scholar to scrub through its talk video to search for a screencast of the system to see it in action.
This disconnect prohibits readers from fluidly transitioning between papers and talk videos because context switching can be disruptive~\cite{chang1998negotiation} and incurs significant cognitive load~\cite{ayres2012split}. As a result, while the research community has recently made significant efforts in creating presentation talk videos and making them widely available even after conferences, researchers are unable to fully capitalize on their benefits.

In a formative study with researchers (n=14), we investigated opportunities and challenges in consuming papers and videos together, and the design space for combining these two formats.
Instead of augmenting one format with the other, our findings revealed that researchers alternated their focus between the paper and video to control the level of detail in which they consumed the paper.
Additionally, researchers observed how linking video segments to relevant paper passages (e.g., paragraphs, figures) could facilitate navigation, as the video could act as a visual map for the paper.
Finally, researchers were against replacing or overlaying content in one format with content from the other as this could obscure information and the effort they dedicated in authoring both formats.

Based on these findings, we designed a novel paper reading experience, \textit{\sysname{}s} (\textbf{pap}er and vid\textbf{eo}), that integrates segments of the talk videos as localized \textit{video notes} alongside corresponding sections of the paper.
As a user scrolls through a \sysname{}, they can see color-coded \textit{highlight bars} in the paper that hint at meaningful passages that have been covered in the video and, next to the paper, correspondingly color-coded video notes with thumbnails of the relevant video segments. 
When the user struggles to understand a portion of the paper, they can click on the highlight bar or video note to play the segment and get a summarized, alternative explanation.
Instead of scrolling through the paper, the user can also choose to focus on the video by navigating between video notes or ``auto-playing'' through them. To avoid disturbing the user's watching, the system fixes the video note's position in the viewport and scrolls the paper to the relevant passage. 
To grant authors control on how \sysname{}s are created for their papers and facilitate the creation process, we also present an authoring interface where authors can link their papers and talk videos with the help of AI suggestions.

To evaluate \sysname{}s, we conducted a within-subjects study (n=16) where participants read and wrote a summary for the systems section of three papers using only the paper, the paper and talk video, or a \sysname{}.
Our study revealed that \sysname{}s could help researchers understand papers and decrease their mental demand during reading. 
Additionally, through \sysname{}s, each format became a guide for the other which facilitated participants' navigation in the two formats and encouraged them to interact with both formats more.
As a consequence of the reduced cognitive demand and improved navigation support, participants composed summaries that more comprehensively covered details from the papers.
In addition, we conducted a field deployment of \sysname{}s during an HCI conference where we had over 250 unique visitors to our reading interface.

This paper presents the following contributions:
\begin{enumerate}
    \item A formative study using a design probe (Fig.~\ref{fig:probe}) with 14 participants that revealed user needs and potential benefits of combining talk videos and research papers for readers.
    \item Co-design sessions with 14 paper authors that focused on understanding how authors would like to combine their papers and talk videos, to explore the design space for combining scholarly papers with talk videos. 
    \item \sysname{}s: A novel reading experience that augments research papers with margin notes that present segments from a talk video alongside relevant passages in the paper (Fig.~\ref{fig:reader}).
    \item A mixed-initiative authoring interface that facilitates the creation of \sysname{}s through AI-based suggestions, to explore the costs and feasibility of creating \sysname{}s (Fig.~\ref{fig:authoring}).
    \item A within-subjects study with 16 participants that revealed how integrating talk videos into papers enables readers to leverage both formats for improved understanding and navigation.
\end{enumerate}

\section{Related Work}

The goal of this work is to explore the design space for augmenting scientific paper reading with corresponding presentation talk videos. To better understand this space, we first review literature around these formats: tools that support general reading, scholarly reading, and knowledge consumption using videos. Finally, we also review prior techniques in other domains for linking between text documents and videos.

\subsection{Augmented Reading Interfaces}
The advent of computers has enabled the creation of reading environments that transcend the limitations of static print media and, instead, allow knowledge workers to interact with and explore text dynamically~\cite{victor2014humane,victor2011explorable}.
Hypertext~\cite{conklin1987hypertext} interconnected scattered text and documents, and this concept has been widely adopted in many reading tools today (e.g., Amazon Kindle's in-situ definitions~\cite{wordwise}, and Wikipedia's page previews~\cite{mediawiki}).
Expanding on hypertext, fluid documents~\cite{chang1998negotiation} and fluid links~\cite{zellweger1998fluid} restructure documents to incorporate this linked content within the document, and various interfaces provide links between text and other document objects, such as tables~\cite{kim2018facilitating} or visualizations~\cite{badam2018elastic}.
To support active reading, various interfaces allow readers to annotate documents with multiple modalities, such as ink or voice~\cite{schilit1998beyond, yoon2014richreview}, to manipulate the document's structure~\cite{tashman2011liquidtext}, or to ask questions and find answers during reading~\cite{chilana2012lemonaid, head2015tutorons, hullman2018improving}.
As documents are frequently dense in content, researchers have investigated how to scaffold navigation by providing overviews~\cite{graham1999reader, schilit1998beyond}, highlighting or fading out content to direct readers' attention~\cite{kelleher2005stencils, yang2017hitext}, or guiding readers based on the activity of other readers~\cite{hill1992edit, kim2014content}.
Extending on this rich body of work, we investigate how to augment the dynamism of academic papers by leveraging and integrating existing talk videos. 

\subsection{Tools for Reading Scientific Papers}

A variety of tools have been designed to address the challenges in reading papers~\cite{lo2023semantic}.
As a crucial component of reading a paper is to contextualize it within the broader literature, CiteRead~\cite{rachatasumrit2022citeread} augments a paper with commentary from citing papers, CiteSee~\cite{chang2023citesee} contextualizes inline citations to a reader's previous reading and publishing activities with visual augmentations, and Threddy~\cite{kang2022threddy} and Synergi~\cite{kang2023synergi} allow users to clip citing sentences and references to explore related themes and papers in the literature.
More closely related to our work, there is a line of research that focused on enhancing both efficiency and comprehension during paper reading. 
Specifically, to help readers traverse the complex language and notation used in scientific papers, Paper Plain~\cite{august2022paper} provides definitions for unfamiliar terms and in-situ summaries of sections, and ScholarPhi~\cite{head2021augmenting} surfaces position-sensitive definitions for unique terms and symbols. 
Also, to facilitate skimming of papers, Scim~\cite{fok2022scim} highlights salient passages of the paper to direct readers' focus, and Spotlights~\cite{lee2016spotlights} surfaces important objects as temporary overlays to help readers identify them even as they quickly scroll through the paper.
Finally, since most scholarly papers are available as PDFs, various approaches have aimed at overcoming the limitations of this format to increase accessibility~\cite{wang2021improving, park2022exploring} and dynamism (e.g., embedding animations~\cite{grossman2015your} or interactive elements~\cite{masson2020chameleon}).
While prior work have focused on designs that can support specific user needs such as skimming \cite{fok2022scim} or simplification \cite{august2022paper}, in this work, we explore how incorporating talk videos has the potential to embody multiple user needs when reading a paper.
Specifically, a talk video can present an author-curated summary for the paper, highlight significant aspects of the work. Linking video segments back to their corresponding passages in the papers also has the potential of allowing readers to skim the paper based on the passages that the authors selected to include in their talk videos.
Furthermore, talk videos include additional commentary, audibly narrate the content which can supplement screen readers, and dynamically illustrate aspects of the work such as animations and screen recordings.

\subsection{Video-based Knowledge Consumption}

Videos are increasingly becoming a predominant channel through which people consume and learn knowledge.
According to Mayer and Moreno's principles~\cite{mayer1998cognitive}, videos can be cognitively beneficial as verbal and visual explanations allow viewers to build two mental representations~\cite{mayer1992instructive, mayer1991animations} without mental overload as audio and visual channels can be processed simultaneously~\cite{mayer1998split}.
As support to these principles, various studies have demonstrated that videos can benefit learners in various domains~\cite{stockwell2015blended, kay2012exploring, hsin2013short}.
While effective for consumption of knowledge, videos represent a continuous stream of frames, and it can be inherently difficult to skim through or locate information in videos, which prior work had shown to be a common need for scholars~\cite{fok2022scim}.
To overcome this limitation and harness the potential of videos, various tools have been designed to facilitate video navigation in learning contexts~\cite{kim2014crowdsourcing, kim2014data, liu2018conceptscape}.
In this work, we investigate the benefits of talk videos for consumption of research, and how to combine these with papers to support both video and paper navigation---allowing scholars to fluidly switch between the two formats.

\subsection{Bridging Text Documents and Videos}

To overcome the difficulty in skimming and efficiently navigating videos, prior work has investigated various approaches to bridge videos with relevant text documents in a variety of domains.
In education, Video Digests~\cite{pavel2014video} and VideoDoc~\cite{krosnick2015videodoc} segment lecture videos into sections so that students can navigate between different parts of a lecture with transcript summaries, and
Shin et al.~\cite{shin2015visual} further combined transcripts with extracted blackboard notes.
Beyond lecture videos, Truong et al.~\cite{truong2021automatic} transform transcripts into hierarchical tutorials for instructional makeup videos, and Sceneskim~\cite{pavel2015sceneskim} facilitates searching and browsing by temporally aligning movies with their captions, scripts and summaries.
Further, Codemotion~\cite{khandwala2018codemotion} automatically extracts code shown in programming tutorials to allow the user to navigate tutorials based on code-related steps. 
While existing research above focused on improving video navigation with text extracted from the same videos (e.g., audio transcripts or blackboard notes extracted from the frames), in this work, we explore how to bridge talk videos with research papers, which are separate entities and different media, and investigate how combining them can facilitate navigation for both media and help scholars better comprehend prior research.
\section{Formative and Co-Design Study}

To explore the design space for combining research papers and talk videos, we conducted a formative study where participants explored the opportunities and challenges in combining the two formats from the perspectives of both readers and authors.

\subsection{Participants}

We invited 14 researchers who had previously published at least one paper and created accompanying talk videos. 
10 were doctoral students, 2 were Master's students, and the remaining 2 were a postdoc and an undergraduate student. 
10 of the 14 participants identified their discipline as human-computer interaction (HCI) or related sub-fields (e.g., visualizations, AI fairness), 3 as natural language processing (NLP), 2 as machine learning (ML), and 1 as computer vision (CV).\footnote{Several participants identified with multiple disciplines.}

\begin{figure*}[!hb]
    \centering
    \includegraphics[width=0.90\textwidth]{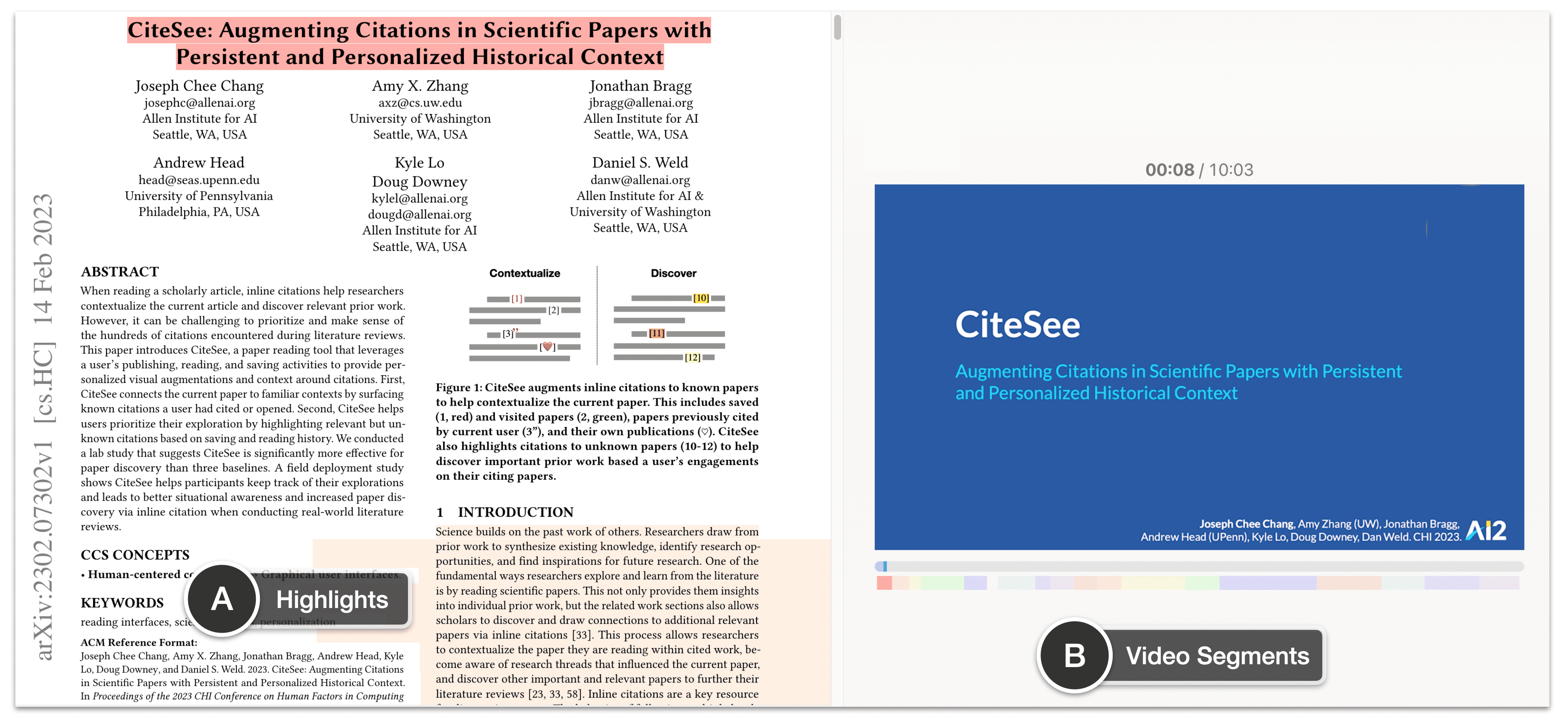}
    \caption{Technology probe used during the formative studies. On the left, a PDF reader for the paper where passages linked to video segments are highlighted (A). On the right, a video player for the talk video accompanied by an interactive timeline and a bar displaying the location and length of segments linked to the paper (B). Linked passage-segment pairs are color-coded.}
    \label{fig:probe}
\end{figure*}

\subsection{Apparatus}

Consuming scholarly papers and talk videos at the same time is a new experience that may be hard for participants to imagine. 
In a preliminary version of this formative study, we gave participants (n=4) a paper and talk video pair side-by-side and instructed them to \textit{``understand the content of the paper based on your real-life habits''}.
Although participants could freely choose how they wished to consume the paper and video, they all watched the whole video first and then delved into the paper.
Participants expressed how this was not due to a lack of desire to jump to the paper while watching the video, but due to the prohibitively high cost of cross-referencing between formats.
This preliminary study revealed that unaugmented papers and videos were inadequate to explore how readers wanted to leverage both formats together.

Thus, we developed a technology probe~\cite{hutchinson2003technology} (Fig.~\ref{fig:probe}) where we could pre-link segments of a talk video to relevant passages in the paper (e.g., paragraphs, figures) and color-code them so that participants could switch between the two formats with lower cost.
Before the study, one of the authors manually created the links between the papers and videos for three papers in each of the recruited participants' research fields (e.g., empirical HCI, systems HCI, NLP, CV).
To create these links, the author followed criteria that were based on insights from the preliminary study: segment the video on slide transitions, and link segments to paragraphs based on content similarity (e.g., phrases, figures) while following the paper's reading order.

\subsection{Study Procedure}

The study consisted of two consecutive sessions.
First, there was a formative session where participants took the perspective of paper readers and used the technology probe (Fig.~\ref{fig:probe}) to read a paper where several passages were pre-linked to relevant segments of the talk video. Then, in a co-design session, participants took the perspective of paper authors and considered designs for combining their own research papers and talk videos. 

For the formative session, participants chose their preferred paper from the set of pre-linked paper-video pairs and, while thinking aloud, read the paper using the technology probe for 20 minutes.
In the probe, linked passages in the paper were highlighted, and participants could click on a linked passage to automatically navigate to the corresponding segment in the video. 
The video segments were also displayed under the video timeline, and participants could click on a video segment to scroll to the corresponding passage in the paper.
After the reading period, participants were asked about the benefits and drawbacks of using the probe and the talk video during paper reading.

Then, participants took the perspective of authors and participated in a co-design session where they considered designs for combining their own research paper and talk video.
To stimulate the participants and illustrate how to sketch designs, participants were provided with a slide deck that showed three example designs for interfaces that combined papers and videos.
Participants were asked to think aloud and sketch designs in the slide deck, which was pre-populated with screenshots of the pages and key frames of the participant's paper and talk video that they provided prior to the study.
To sketch out their designs, participants could resize and crop the screenshots, draw shapes, and use text boxes to describe the designs.
During the session, one or two of the authors helped with the sketching by making edits based on participants' descriptions, and asked questions to encourage participants to elaborate further on their ideas or to consider alternative designs. 

Aside from one in-person participant, all participants joined remotely through Google Meet.\footnote{\url{https://meet.google.com}} This study was approved by our internal review board, and each participant was paid 45 USD for their time. 

\subsection{Findings}

During the study sessions, we recorded participants' screens and the audio, which were then manually transcribed.
Through a thematic analysis, the transcripts were coded and these codes were grouped into themes to identify the main insights from the study.
Additionally, a thematic analysis was also conducted on the various designs for the co-design sessions to typify these designs based on their similarities.
Based on insights from the reader and author sessions, we distilled design goals for augmenting research papers with talk videos.

\subsubsection{As Readers}

In contrast to participants in the preliminary study, participants in this study followed different consumption patterns with the probe: five mainly read the paper and occasionally switched to the video, and nine followed the video while intermittently pausing to dive into the paper.
Based on their experiences with the technology probe, participants noted various ways in which talk videos enriched the paper.
Specifically, most participants (11/14) mentioned that the video provided summaries that were easier to consume than \textit{``dense parts of the paper''} (P5).
Asides from summarizing, participants (7/14) also mentioned that videos explain details differently and that these alternative explanations were useful when they struggled to understand the paper.
Participants also noted the significance of the audio-visual nature of videos.
Several participants liked authors' narrations in videos (4/14) as listening could be less demanding or more \textit{``passive''} than reading (P10), and since they could have the \textit{``author narrate [figures] for [them]''} (P2).
In terms of the visuals, various participants (5/14) described how illustrations, animations, or clips in the talk videos could better illustrate certain aspects of the paper. For example, P14 mentioned how a clip showing a demo of an interface helped them \textit{``get a more clear idea of what the interaction would look like''}.
Finally, a majority of participants (11/14) mentioned how watching the videos or skimming the video-based highlights in the paper gave them an overview of the papers and allowed them to \textit{``make note''} (P1) of details they wanted to dive deeper into---serving as a \textit{``launching pad''} into the paper (P3).

\begin{table*}[!t]
\begin{tabular}{@{}cll@{}}
\toprule
\textbf{Primary Format} &
  \textbf{Type of Design} &
  \textbf{Feature Differences} \\ \midrule
\multirow{11}{*}{\textbf{Paper}} &
  \multirow{3}{*}{\begin{tabular}[c]{@{}l@{}}\textbf{Linked video popups}: display popup\\ with video segment when user \\ interacts with a linked paper passages.\end{tabular}} &
  \begin{tabular}[c]{@{}l@{}}Link popups on text (P4, P8, P11, P13), figures or tables (P2, P6, \\ P8), or definitions and sections headers (P14).\end{tabular} \\ \cmidrule(l){3-3} 
 &
   &
  Display popup based on user’s selected text (P3). \\ \cmidrule(l){3-3} 
 &
   &
  Display thumbnail instead of video segment (P6, P7). \\ \cmidrule(l){2-3} 
 &
  \multirow{3}{*}{\begin{tabular}[c]{@{}l@{}}\textbf{Overlaid videos}: overlaying video \\ segments on relevant passages of \\ the paper.\end{tabular}} &
  Overlay on videos on figures (P4, P8, P13). \\ \cmidrule(l){3-3} 
 &
   &
  \begin{tabular}[c]{@{}l@{}}Overlay visual guides from video on tables or figures (P2, P6, P8), \\ or mathematical equations (P8).\end{tabular} \\ \cmidrule(l){2-3} 
 &
  \multirow{4}{*}{\begin{tabular}[c]{@{}l@{}}\textbf{Video-based outline}: an outline or \\ table of contents for the paper based \\ on the links between video segments \\ and paper passages. \end{tabular}} &
  \begin{tabular}[c]{@{}l@{}}Panel that displays a list of the slides extracted from the video as a \\ navigational map (P2).\end{tabular} \\ \cmidrule(l){3-3} 
 &
   &
  \begin{tabular}[c]{@{}l@{}}Table of content for the paper but containing the titles of video \\ sections (P11), and transcript summaries or video thumbnails (P12).\end{tabular} \\ \midrule
\multirow{8}{*}{\textbf{Video}} &
  \begin{tabular}[c]{@{}l@{}}\textbf{Position-sensitive details}: hovering \\ over elements in a video frame to \\ reveal a tooltip with related details \\ from the paper.\end{tabular} &
  \begin{tabular}[c]{@{}l@{}}Hovering over keywords to see definitions (P7, P10), summarized \\ tables to reveal the detailed tables from the paper (P10), or \\ elements of a system to reveal related explanations from the \\ paper (P13).\end{tabular} \\ \cmidrule(l){2-3} 
 &
  \begin{tabular}[c]{@{}l@{}}\textbf{Guiding tooltips}: tooltips that appear as\\ the video plays to encourage the viewer \\ to check related sections of the paper.\end{tabular} &
  \begin{tabular}[c]{@{}l@{}}Tooltip is accessible through an icon that is overlaid on the video \\ (P1), tooltip text is overlaid on the video (P10, P14) or text is \\ shown next to the video (P1).\end{tabular} \\ \cmidrule(l){2-3} 
 &
  \begin{tabular}[c]{@{}l@{}}\textbf{Side commentary}: panel next to the \\ video that displays relevant passages \\ from the paper as the video plays.\end{tabular} &
  \begin{tabular}[c]{@{}l@{}}Commentary can include the full passages from the paper (P13), \\ only information from the passages that is not included in the \\ video (P5), or a summary of the passages (P5, P13)\end{tabular} \\ \midrule
\multirow{4}{*}{\textbf{Combined}} &
  \begin{tabular}[c]{@{}l@{}}\textbf{Interweaved paper and video}: new \\ format that interweaves elements from \\ the paper with those from the video.\end{tabular} &
  \begin{tabular}[c]{@{}l@{}}Embedding images, animated GIFs, and clips from the video \\ inbetween passages of text (P6), inbetween summarized passages \\ of text (P3), or replace text with the video elements (P2, P9).\end{tabular} \\ \cmidrule(l){2-3} 
 &
  \begin{tabular}[c]{@{}l@{}}\textbf{Adaptive side-by-side}: paper and video \\ displayed side-by-side but adaptively \\ changes the size of each format.\end{tabular} &
  \begin{tabular}[c]{@{}l@{}}User can manually change the amount of space taken by each \\ format or the interfaces automatically changes them by inferring \\ the user’s needs (P4).\end{tabular} \\ \bottomrule
\end{tabular}
\caption{Overview of the co-design session that captured how authors envisioned combining their papers and talk videos. The table describes the types of designs that authors produced and the features that authors proposed for the different design types. Additionally, the design types were categorized based on their primary consumption formats.}
\label{tab:author-designs}
\end{table*}

Despite these benefits, however, there were various interaction challenges that limited participants' use of talk videos even with the support of our technology probe (Fig~\ref{fig:probe}).
For example, as a paper automatically scrolled to the relevant passage when the video progressed to the next segment, participants mentioned that the probe could disrupt their reading (3/14) or cause them to get lost (6/14).
Additionally, participants (4/14) mentioned how they could not predict what information would be contained in a video segment before actually watching the segment and, therefore, could not anticipate when a segment would be useful or not.
Finally, as video segments were linked to relatively lengthy passages in papers, various participants (8/14) mentioned how it was difficult to locate a detail mentioned in a video segment in the paper, or to distinguish what in the paper passages had been covered or not by the segment.

\subsubsection{As Authors}

During the co-design sessions, participants produced a variety of designs for paper and video combinations. 
As seen in \autoref{tab:author-designs}, several of the participants' designs shared structural similarities, but differed in terms of specific details or features.
Participants considered both designs where the video supported paper reading and where the paper enhanced video watching, and some participants envisioned new formats where neither format was the main one.

Based on participants' designs and their comments during the sessions, we distilled the main goals that participants considered when designing the combinations. 
One of the main goals that participants (12/14) mentioned was to enable users to flexibly switch the level of detail at which they consume the content. 
Specifically, the user can switch from the video to the paper to \textit{``expand to see more details''} (P1) or switch from the paper to the video to \textit{``skip''} (P5) sections that are less interesting. 
Beyond consumption, several participants (7/14) considered combinations that visually represented paper passages with the video to support navigation in the paper. 
For example, P2's design presented slides from the video as a visual outline that the user can use to navigate the paper.
Finally, due to their difficulties in locating details from the video in the paper and vice-versa during the reading session, several participants (5/14) designed interfaces that supported more fine-grained links (e.g., highlighting passages in the paper that were mentioned in the video).

Beyond revealing what authors wanted from the combinations, the co-design sessions also revealed constraints to possible designs.
While several participants created designs that replaced paper passages with video elements, most participants (7/14) advocated against replacing content. 
Some participants mentioned that \textit{``videos are rarely a one-to-one representation of a paper''} (P3) and that replacing could \textit{``delete information''} (P10), while others noted how one format provided \textit{``supplementary information''} for the other (P6) and it could be more beneficial to consume them together.
Additionally, P2 mentioned how they dedicated \textit{``significant effort''} in authoring their paper and video, and that they would want users to look at both artifacts.
Another constraint was that, despite considering designs where the user mainly interacted with the video, most participants (7/14) considered the video as \textit{``a way to advertise''} their paper (P4) and that \textit{``ultimately''} (P11) they wanted to direct the user to their paper. This was reflected through their \textit{``guiding tooltip''} designs (\autoref{tab:author-designs}).

Finally, we asked participants about whether they would be willing to create links between their papers and talk videos to enable the combinations they designed. 
All participants mentioned that they would create these links as it could increase the visibility of their work and \textit{``help as many people as possible to read and understand [my paper]''} (P9).
Although several participants mentioned that they would want the process of linking to be as easy as possible, all participants also mentioned that they would not want the process to be completely automatic.
Instead, they would need to be \textit{``involved in the process''} (P3) to check and edit links made by an automatic pipeline.
Interestingly, some participants even expressed how they would be willing to change how they author videos to make this semi-automatic linking process easier and more accurate: \textit{``I might start baking this stuff into the slide deck''} (P10) and \textit{``It might have a positive influence on [...] how I design the the slides like making them more correlated to the paper''} (P6).

\subsubsection{Design Goals}

Based on the insights from the reader and author sessions, we distilled the following design goals for combining research papers and talk videos:

\begin{itemize}
    \item DG1: Allow readers to both focus on either the paper or video, but also enable them to fluidly switch between the two formats when needed. 
    \item DG2: Surface visuals from the video to help readers anticipate its content and to visually outline the paper.
    \item DG3: Present fine-grained links that aid in the association of related details across formats.
    \item DG4: Avoid occluding or replacing the content in a format with content from the other.
    \item DG5: Aid in the creation of links between papers and videos but grant authors control over how they want to present their work.
\end{itemize}

\begin{figure*}[!b]
    \centering
    \includegraphics[width=1.0\textwidth]{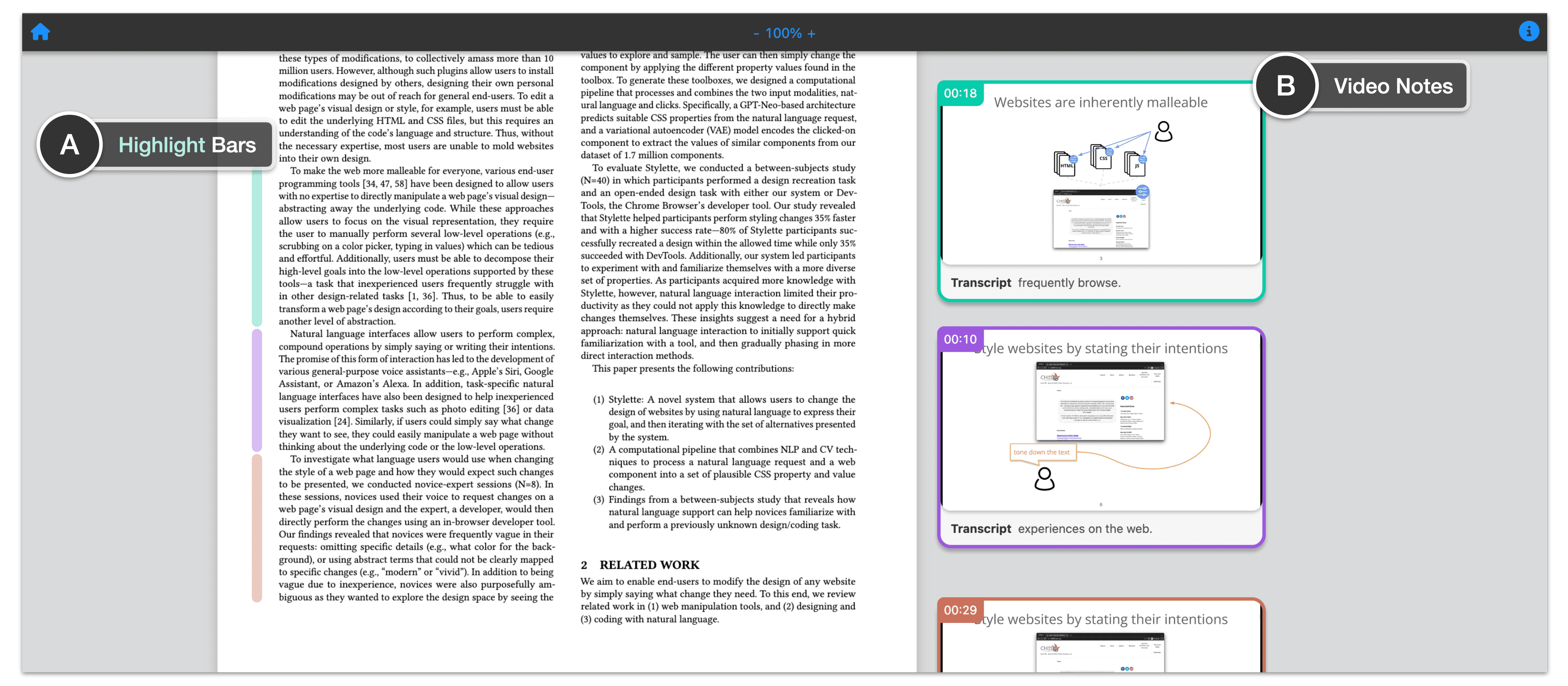}
    \caption{The \sysname{} reader extends a PDF reader by incorporating highlight bars (A) alongside passages in a research paper that are linked to segments in the corresponding talk video. These video segments are displayed as video notes (B) that are localized next to the linked passages and present a thumbnail, a line from the transcript, and the total duration of the segment.}
    \label{fig:reader}
\end{figure*}

\section{Papeos}

Based on the design goals, we developed \textit{\sysname{}s} (\autoref{fig:reader}), a novel reading experience that augments research papers with localized clips from the corresponding talk videos.
In this section, we first illustrate the reading interface for \sysname{}s.
Then, we describe a mixed-initiative interface that allows paper authors to create \sysname{}s for their papers and talk videos with lowered effort.

\subsection{Papeo Reading Interface}

The \sysname{} reader is designed to support a variety of use cases, such as leveraging linked video segments to guide users when text skimming (\S\ref{sec:skim}), support users in fluidly switching between reading text passages and watching video segments to adjust the level of details they wish to consume (\S\ref{sec:fluid}), and allow users to continuously watch a talk video while having access to additional details in corresponding text passages (\S\ref{sec:video}).
For this, the \sysname{} reader presents video segments as \emph{video notes} placed on the right side of paper pages and localized approximately next to their linked passages (Fig.~\ref{fig:reader}).
Since each page of a paper could contain multiple linked passages and video notes, \sysname{} renders color-coded \emph{highlight bars} next to passages and video notes alongside a paper for linked paper passages and video segments (DG4).

\subsubsection{Video-Supported Skimming}
\label{sec:skim}

\begin{figure*}[!t]
    \centering
    \includegraphics[width=1.0\textwidth]{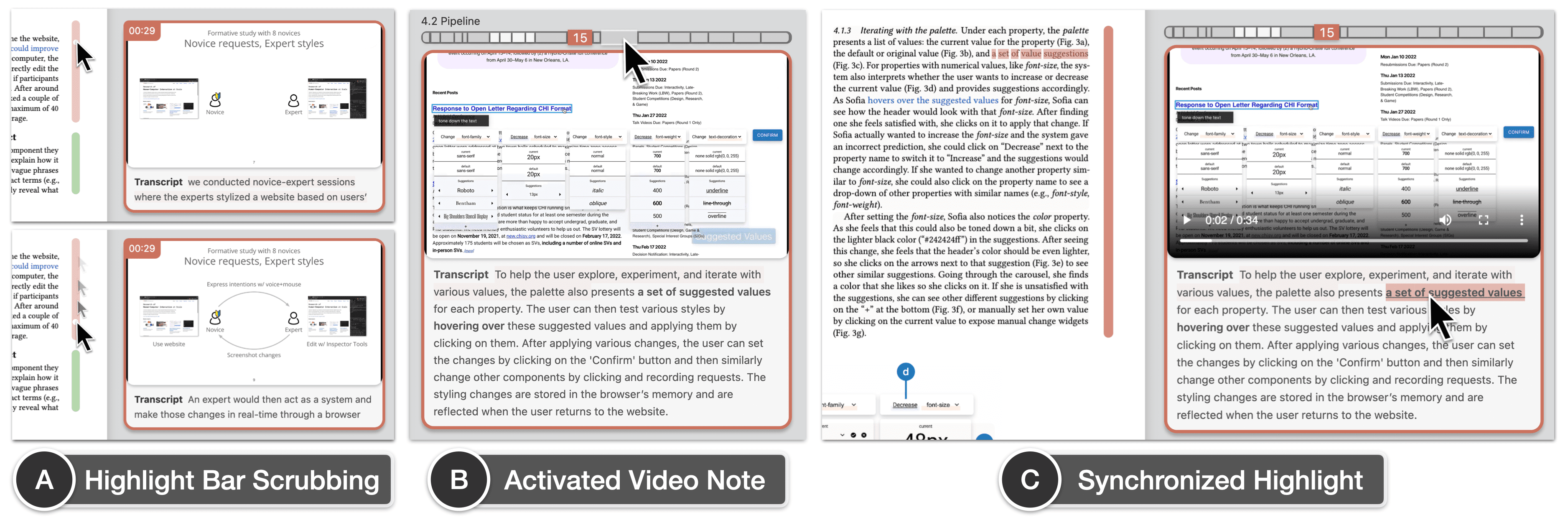}
    \caption{Illustration of features supported by the \sysname{} reader: (A) hovering and scrubbing over highlight bars allows users to quickly scrub through the linked video segments; (B) activated video notes present the users with player controls, the full transcript for the segment, and a segmented timeline for the whole video that presents the paper section where a note is located when the user hovers a segment; and (C) synchronized highlights are shown as blue text in the paper and bold text in the video transcript, and, when the user hovers over them, they become highlighted in sync.}
    \label{fig:reader-features}
\end{figure*}

Researchers often skim read to get a high level understanding of research papers \cite{fok2022scim}. By scrolling through the \sysname{} reader, the user can skim the paper by looking through the highlight bars and accompanying video notes. 
The highlight bars (Fig.~\ref{fig:reader}a) reveal the portions of the paper that the author considered important when creating their video.
The video notes (Fig.~\ref{fig:reader}b) reveal the content of the video segment through the thumbnail (i.e., the first frame of the video segment) and the first line from the transcript which can, respectively, visually represent and summarize these passages of importance.
By skimming based on these features, for example, a reader could prioritize reading high-level descriptions of a user interface and a few important quotes from the user study that were included in the conference presentation, instead of reading all implementation details and quotes that were not included.
By remembering the thumbnails and their relevant locations in the paper, the user can also develop a ``spatial mental map''~\cite{rachatasumrit2022citeread} of the paper to help them return to desired content in the paper (DG2).
If the thumbnail or transcript line surfaces insufficient information about the video segment, the user can also hover and scrub over the highlight bar to peek into different moments in the segment (Fig.~\ref{fig:reader-features}a). 

\subsubsection{Fluid Switching between Paper and Video}
\label{sec:fluid}

As the user is reading through the paper, they may struggle to understand certain passages or may be less interested in particular sections. For example, an expert user might need to learn the implementation details of a machine learning paper but was already familiar with the background and related work.
In these cases, if a video note is linked, the user can watch an alternative and/or summarized explanation of the passage by clicking on the highlight bar or video note itself (DG1). 
Clicking on the bar or note ``activates'' the video note (Fig.~\ref{fig:reader-features}b): the thumbnail switches into a video player that starts playing the segment, the full transcript for the segment is shown, and the note increases in size.
If it is only approximately aligned with the highlight bar, the note also moves to be exactly aligned---pushing away other notes if they would overlap. 
As the video plays, lines of the transcript are highlighted so that the user can discern what has already been spoken.

While watching the video note, the user may want to read up on the same information in the paper to acquire more details or to take in a more formalized explanation.
To focus back on the reading, the user can pause the video note through the player controls or by clicking anywhere outside the note to ``deactivate'' it.
As users may start reading while the video note plays and forget to deactivate it, each video note only streams one video segment to minimize disruption.
Thus, by default, once the video note reaches the end of the current segment, the player stops instead of progressing to the next segment in the video---unlike the preliminary research probe (DG1).

\begin{figure}[!b]
    \centering
    \includegraphics[width=1.0\columnwidth]{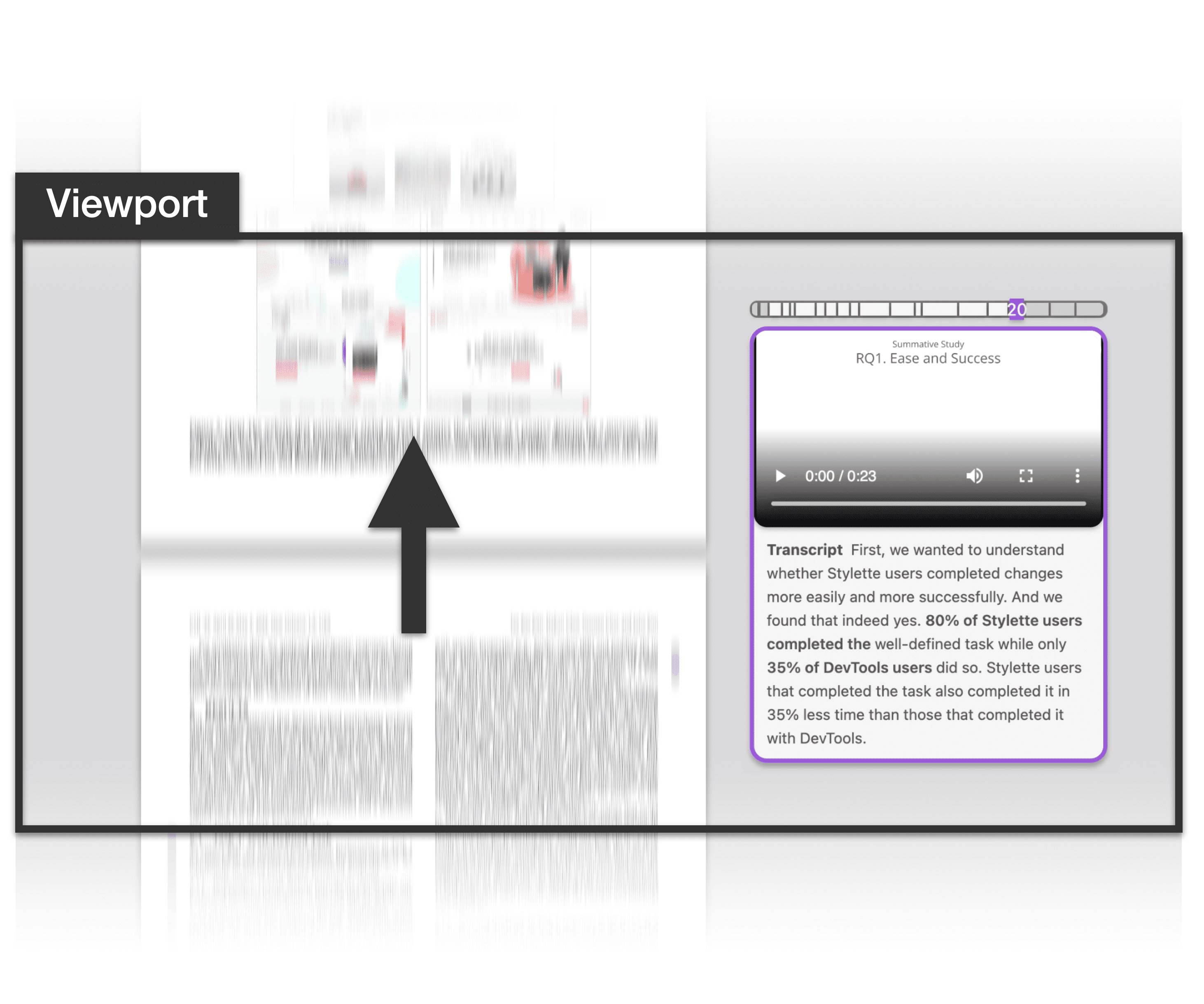}
    \caption{During video note-centric scrolling, the user can navigate to the video note for the next video segment, which takes over the viewport position of the current video note. With the video note fixed in position, the paper scrolls to the passages linked to the next video segment. This allows the user to continuously watch video segments without interruption while always having access to the linked passages next to the current video playback.}
    \label{fig:scrolling}
\end{figure}

Finally, when switching between the two formats mid-segment, the user may struggle to identify a detail in one format in the other due to the wording differences or the amount of text they have to traverse through. 
For example, if a reader watches a progressive animation explaining the architecture of a machine learning model and becomes curious about a specific hyper-parameter, it can be difficult for them to find the value of the hyper-parameter in the paper.
To address this challenge, the \sysname{} reader provides \textit{synchronized highlights} (Fig.~\ref{fig:reader-features}c).
Based on how the paper author created the \sysname{}, certain words or phrases in the video transcript and paper are bold and underlined.
When the user hovers over these words or phrases, they are highlighted and the related words or phrases in the other format are also highlighted to help the user discern and match details across the formats (DG3).

\subsubsection{Video-Centric Consumption}
\label{sec:video}
Besides skimming the text of the paper and switching between text and video segments, \sysname{} also support users if they wish to watch multiple segments or even the entire video continuously. While each video note only streams one segment from the video, the \sysname{} reader also allows the user to focus on and watch the video notes in order (DG1).
When a video note ends, the user is provided with the option to re-watch the video segment or to jump to the next. 
To watch the whole video with no interruptions, the user can activate the ``autoplay'' setting to automatically navigate and watch through all video segments.

Whenever the user navigates between video notes, the paper scrolls automatically to the location of the next video note to allow the user to also check and read the linked paper passages (DG1).
To minimize disruption during autoplay, the \sysname{} reader employs \textit{video note-centric scrolling} (Fig.~\ref{fig:scrolling}).
In this type of scrolling, the different video notes stay fixed in same position while the paper scrolls to corresponding linked passages as the videos play.
Thus, while the user is technically navigating between video notes and scrolling through the paper, they can continue to watch the video by fixing their gaze on the same part of their screen (DG1). 

Above activated video notes, the \sysname{} reader also provides a timeline (top in Fig.~\ref{fig:reader-features}b) to allow the user to navigate between video notes and, consequently, navigate the paper based on these (DG2). 
The timeline is fragmented where each block represents a video note and the user can navigate to these notes by clicking on the blocks---navigation occurs through \textit{note-centric scrolling}.
To help the user track where they are in the video and what they have already watched, the block for the current video note is color-coded and blocks for notes that have been watched are opaque.
Before navigating to a note, the user can hover over a block to see the title of the section or sub-section where the note is located (\textit{``4.2 Pipeline''} in Fig.~\ref{fig:reader-features}b)---allowing them to check where they will navigate to and what type of content may be contained in the video note (DG2).

\begin{figure*}[!b]
    \centering
    \includegraphics[width=1.0\textwidth]{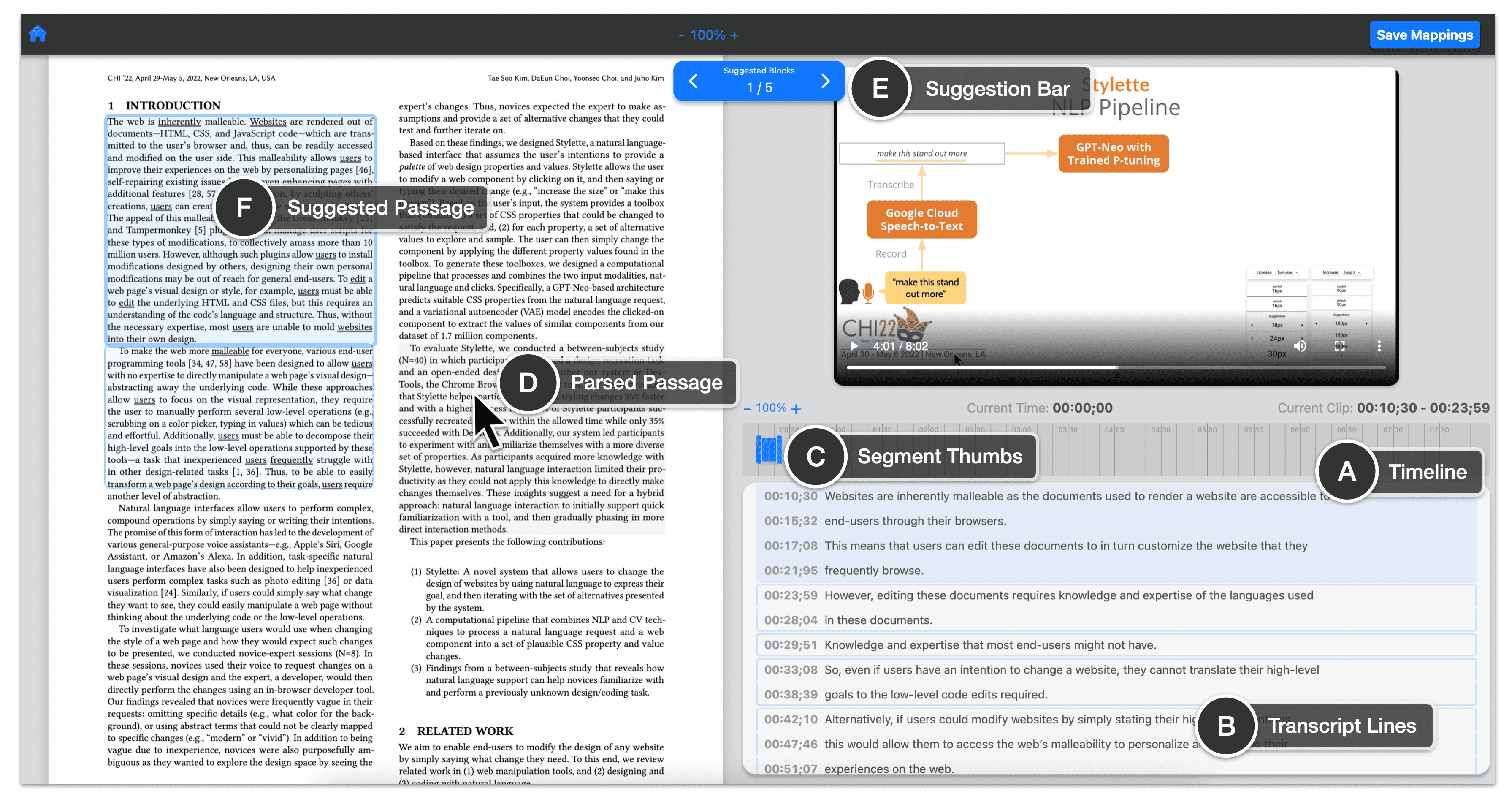}
    \caption{The \sysname{} authoring interface consists of a parsed PDF and a video segmenter. The segmenter timeline (A) displays the segments that have been created so far. The user can create a segment by clicking on the timeline or selecting lines in the transcript (B), and then dragging on the thumbs to fine-tune its length (C). Then, the user can manually click on relevant passages (D) to link them to a video segment, or review and adopt the automatically generated linking suggestions (E,F).}
    \label{fig:authoring}
\end{figure*}

\subsection{Papeo Authoring Interface}

To create \sysname{}s, we propose an authoring interface (\autoref{fig:authoring}) through which paper authors can link their papers and talk videos---granting them control over how these formats are linked (DG5).
We developed this interface through an iterative design process. 
With early versions of the interface, we observed that authors dedicated significant effort to segment their videos and to search for paper passages that were relevant to these segments.
To address this challenge, we adopted a mixed-initiative design for the authoring interface by providing automatic suggestions for segmenting videos and for linking papers and videos.
To start authoring, the author first uploads a PDF of their paper and the talk video with transcript. 
Then, they access the authoring interface that consists of two panels: a video segmenter where the author can select segments of the video, and a paper annotator where they can then choose the passages to link to the segment (Fig.~\ref{fig:authoring}).

To start linking their paper and video, the author first needs to create a video segment. To do so, they can watch the video, click on the timeline to create an initial segment, and drag the start and end thumbs to select a time range (Fig.~\ref{fig:authoring}a). Alternatively, authors can read the transcript and directly select a group of transcript lines (Fig.~\ref{fig:authoring}b).
To improve efficiency, the interface also automatically groups transcript lines at the sentence-level to act as segment suggestions.
When the author clicks on a group, the interface selects a segment that contains all of the lines in the group.
When creating a segment from the transcript, the author can select or de-select lines to correct errors in the segment suggestions, or further fine-tune the start and end times by using the thumbs in the timeline since transcript lines do not always align with sentence boundaries (Fig.~\ref{fig:authoring}c).

After creating a video segment, the author can then link it to relevant passages (e.g., paragraphs, figures) in the paper. 
Instead of requiring authors to manually select paragraphs or figures, the interface presents these as clickable targets so that authors can select entire paragraphs with single clicks (Fig.~\ref{fig:authoring}d).
This is made possible by leveraging the pre-trained VILA model to automatically parse the paper PDF and identify paragraph, figure, and table boundaries~\cite{shen2022vila}.
Since AI models can occasionally make mistakes, the author can also click-and-drag over an area of the paper to manually select a passage to recover from errors.
One remaining challenge here is that it can be time consuming to search through the paper for relevant passages. For this, the interface suggests the five most likely passages based on the current video segment (i.e., link suggestions). After a video segment is created, the paper automatically scrolls to the highlighted top-1 suggestion for the author to review (Fig.~\ref{fig:authoring}f). The author can further review the top 2--5 suggestions using the suggestion navigation bar (Fig.~\ref{fig:authoring}e). 

Beyond the coarse-grained links between paper passages and video segments, the \sysname{} reader interface also supports fine-grained links (i.e., synchronized highlights) to help readers identify specific details. 
To create these fine-grained links, authors can select a paper passage or video segment that has been linked and click on the ``Create Sync Highlight'' button at the top of the interface.
Then, the author can select words or phrases in the passages and the transcript of the video segment that they wish to link.
After selecting the words, the author stores the synchronized highlight by clicking on the ``Save Sync Highlight'' button, and can proceed to create more synchronized highlights for the linked segment and passages.

\subsubsection{Automatic Suggestions}

To make authoring \sysname{}s more efficient, \sysname{}s' authoring interface generates automatic suggestions for video segmentation and for paper-video linking.
During development, to evaluate multiple algorithms and AI models for generating suggestions, we collected a small ground-truth dataset by having three of the authors and five recruited researchers link their papers and talk videos (total of 8 pairs) using an initial version of the authoring interface without automatic suggestions. 
For techniques with no tunable hyperparameters, we evaluated the technique on the whole ground-truth dataset.
For techniques with hyperparameters, we performed 4-fold cross-validation where 25\% of the data was used to identify the best hyperparameter values and the remaining 75\% was used to evaluate the technique with the best identified hyperparameter values.
For each technique, we specify the hyperparamters, if any.

\textbf{Segment Suggestions}: We tested three different techniques for automatically segmenting videos (i.e., shot detection): (1) calculating pixel changes in the HSV (i.e., Hue, Saturation, and Value) colorspace between adjacent frames~\cite{pyscenedetect}, (2) template matching which calculates the spatial similarity between a frame and the previous key frame~\cite{swanberg1993knowledge}, and (3) segmenting the video at every transcript line containing a punctuation---as authors are likely to transition between scenes at the end of sentences.
Both the HSV change and template matching techniques had two hyperparameters: minimum length of a segment, and threshold (i.e., HSV change or spatial similarity value that needs to be exceeded to predict a segment boundary).
For evaluation, we calculated the number of predicted segment boundaries that were within 3 seconds of ground-truth boundaries to calculate precision, recall and F1-score.
Based on the interaction we designed, we expected that it would be easier (i.e., fewer clicks) for authors to merge segment suggestions compared to splitting them, so we used the F3-score, which gives more weight to favor over-segmenting (i.e., more segments) and decided to adopt the punctuation-based auto-segmenter (\autoref{tab:segmentation}).

\begin{table}[!t]
\begin{tabular}{@{}lccccc@{}}
\toprule
\textbf{Algorithm} & \textbf{Precision} & \textbf{Recall} & \textbf{F1} & \textbf{F2} & \textbf{F3}  \\ \midrule
\textcolor[HTML]{1a4595}{Punctuation}        & \textcolor[HTML]{1a4595}{0.405}           & \textcolor[HTML]{1a4595}{\textbf{0.906}}     & \textcolor[HTML]{1a4595}{0.541}    &  \textcolor[HTML]{1a4595}{0.701} &  \textcolor[HTML]{1a4595}{\textbf{0.786}}      \\
HSV Change & 0.499           & 0.805     & 0.605  & \textbf{0.706} & 0.751         \\
Template Match  & \textbf{0.577}  & 0.758   & \textbf{0.635} & 0.698 & 0.725 \\ \bottomrule
\end{tabular}
\caption{Recall, precision, and F1-, F2- and F3-scores for the algorithms tested for video segmentation. Highest values for each metric are shown in bold, and the technique used in the authoring interface is shown in blue.}
\label{tab:segmentation}
\end{table}

\textbf{Linking Suggestions}: Currently, the authoring interface provides \textit{text} passage linking suggestions that appear immediately \textit{after} a video segment was created.
We initially aimed to also automatically identify video frames similar to figures and tables in the papers since authors often reuse figures and tables in their presentations. However, it became clear in early design iterations that mapping figures between papers and videos was a relatively simple task for test users. In contrast, they spent much greater effort when trying to find the right passages when mapping to text.

Therefore, we focused on providing text passage linking suggestions, and used the ground-truth video segments from our dataset to test the following measures for matching text from the segments' transcripts to text in paper passages: (1) cosine similarity based on two text embedding models (i.e., Specter~\cite{cohan2020specter} and MiniLM~\cite{wang2020minilm}), (2) ROUGE-L score~\cite{lin2004rouge}, and (3) a baseline that chooses the first paragraph of a random section in the paper.
We designed the baseline based on the assumption that talk videos provide an overview of the paper and, as a result, might state information included in the overviews of each section (i.e., the first paragraphs).
As seen from the results (\autoref{tab:linking}), ROUGE-L had the highest top-1 accuracy while MiniLM embeddings had the highest top-5 accuracy.
We then combined these two measures by simply adding the two scores which achieved higher top-1 and top-5 accuracies.

Finally, we noticed how videos typically present information content in the same order as the paper---i.e., after linking a segment and passage, the next video segment would likely link to passages that appear later in the paper.
Based on this, we developed an additional technique that adapts the Viterbi algorithm~\cite{forney1973viterbi}.
Using this technique, we can consider, simultaneously, the semantic similarity between paper text and video transcripts, and how information might be presented in similar ordering in the two formats (e.g., background, methods, and then evaluation).
More specifically, the potential links between segments and passages are considered to be states, and an observation is whether the segment and passage are actually linked. In this context, we first normalized the combined measure of MiniLM + ROUGE to use as the emission probability (i.e., probability of linking a segment to each passage). 
Then, we modeled the transition probability as a hyperparameter of the likelihood of linking a video segment to a passage in-order and the remaining probability becomes the likelihood of linking in reverse order.\footnote{Based on the 4-fold cross-validation and grid-search, the transition probability was set to 0.7, 0.5, 0.6, and 0.6 in each fold, respectively.}
This technique improved on both the top-1 and top-5 accuracies and was used to provide suggestions in the authoring interface.

\begin{table}[t]
\begin{tabular}{@{}lcc@{}}
\toprule
\textbf{Algorithm}                  & \textbf{Top-1} & \textbf{Top-5} \\ \midrule
Random first paragraph of a section & 0.029          & 0.080          \\
SPECTER Embeddings                  & 0.399          & 0.623          \\
MiniLM Embeddings                   & 0.464          & 0.768          \\
ROUGE-L Score                       & 0.493          & 0.739          \\
Combined (MiniLM + ROUGE-L)         & 0.572          & 0.797          \\
\textcolor[HTML]{1a4595}{Viterbi with Combined}               & \textcolor[HTML]{1a4595}{\textbf{0.626}} & \textcolor[HTML]{1a4595}{\textbf{0.863}} \\ \bottomrule
\end{tabular}
\caption{Top-1 and top-5 accuracy for the algorithms tested for linking paper passages and video segments. Highest values for each metric are shown in bold, and the technique used in the authoring interface is shown in blue.}
\label{tab:linking}
\end{table}

\subsubsection{Preliminary User Evaluation}
\label{sec:authorUIEval}

To test the feasibility and costs of authors creating Papeos for their readers, we conducted a preliminary evaluation with 6 researchers (3 systems HCI, 3 empirical HCI, and 1 computer vision) to author a \sysname{} using their own research paper and talk videos.
In general, participants mentioned that it was easy to use the authoring interface to link their papers and videos, and that they were enthusiastic to author \sysname{}s for future papers through the interface.
This evaluation demonstrated that participants spent an average of 25 minutes and 22 seconds (SD=5:31, max=30:17, min=15:19) to fully link their paper and video\footnote{In contrast, the five researchers recruited to create the \sysname{} test set, who used the authoring tool without suggestions, took an average of 44 minutes (SD=10.9) to author one \sysname{}.}.
Additionally, we measured how frequently the authors used at least one of the top-5 suggestions when linking a segment to passages, and saw that suggestions were used for 71.3\% (SD=11.6\%, max=82.1\%, min=57.1\%) of all linked segments.
In sum, we showed that authors can use our current authoring interface to create \sysname{}s for their own papers with reasonable effort, and leave further automation and evaluation for future work.

\subsection{Implementation Details}

We implemented the reading and authoring interfaces for \sysname{}s in around 6,500 lines of TypeScript, ReactJS, and CSS.
For the PDF reader, we adapted our own open-source PDF reader library\footnote{\url{https://github.com/allenai/pdf-component-library}} and, for the video player, we used the ReactPlayer package.\footnote{\url{https://github.com/cookpete/react-player}}
The backend and AI-based suggestions were implemented using around 1,600 lines of Python code. We used a Flask server, the HuggingFace Transformer library\footnote{\url{https://huggingface.co/docs/transformers/index}} for the SPECTER \cite{cohan2020specter} and MiniLM 
 \cite{wang2020minilm} models, and the PySceneDetect\footnote{\url{https://scenedetect.com/en/latest/}} and OpenCV\footnote{\url{https://docs.opencv.org/4.x/index.html}} packages for shot detection based on the HSV colorspace and template matching, respectively. 
\section{User Study}

Through our formative study, we observed that talk videos and papers provided different benefits to users. Specifically, we found evidence that talk videos have the potential of complementing paper reading so that the reader can quickly get an overview but also selectively dive deeper into details. However, the interaction cost of fluidly consuming the two formats together can be prohibitively high, which led to a set of design goals that drove the development of \sysname{}.
Thus, we conducted a within-subjects study to investigate whether \sysname{}s can help readers to both acquire a comprehensive understanding of the paper and efficiently identify relevant details.
We compared three conditions: \sysname{}s with linked papers and videos, separated papers and talk videos, and papers only.
With each condition, participants were asked to read the systems section of an assigned paper and to write a summary for the section that was \textit{comprehensive} and \textit{detailed}.
Through this task, we investigated the following research questions:

\begin{itemize}
    \item RQ1. Can \sysname{}s reduce the cognitive load involved in reading and understanding research papers?
    \item RQ2. How do \sysname{}s affect researchers' navigation of research papers and talk videos?
    \item RQ3. Can \sysname{}s help researchers to both comprehensively cover significant aspects of papers and read these in detail?
\end{itemize}

\subsection{Study Design}

\subsubsection{Participants}

We recruited 16 early-stage researchers in HCI for the study through the authors' social media (Twitter) and snowball sampling. 
12 of the participants were first to third year doctoral students, and 4 were Master's students.
Our study focused on early-stage researchers as they may receive the greatest benefit from augmenting paper reading with talk videos---e.g., simplify and visually represent complex explanations, highlight important aspects of a paper.
All participants reported reading research papers at least once a week to several times a day.
The study lasted a total of 90 minutes, and participants were compensated 45 USD for their time. The study was approved by our internal review broad.

\subsubsection{Conditions}

During the study, participants read and wrote summaries for three different papers.
For each paper, they used one of the following conditions: \sysname{}, Paper+Video, and only Paper. 
The ordering of the conditions was counterbalanced to mitigate the influence of ordering effects.
In the \sysname{} condition, the participants used the \sysname{} reader. 
In the Paper+Video condition, participants used a basic PDF reader and a basic video player in separate tabs or windows, and, in the Paper condition, they only used the PDF reader.
The basic PDF reader and video player were developed using the same base libraries and packages as the \sysname{} reader, and provided all basic functionalities available in other similar readers and players (e.g., zoom in, zoom out, playback speed controls).

\subsubsection{Reading Materials}
All of the participants read the same three papers~\cite{kim2022stylette, chang2021tabsdo, huh2022cocomix} in the same order.
We chose the papers from the initial dataset of linked papers and video used to evaluate the automatic suggestions (\S\ref{sec:authorUIEval}).
Specifically, we chose HCI papers that presented systems that incorporated AI or algorithmic pipelines, and whose ``System'' sections were of relatively similar length.
We focused on systems papers as they present interfaces that may be easier to understand with videos.
Additionally, as our goal was to evaluate whether \sysname{}s can help readers identify details, we narrowed down to systems that incorporated pipelines as they may include a substantial amount of design and implementation details.
To match these criteria, we chose two papers written by authors of this paper.
In Appendix~\ref{appendix:papeo-analysis}, we provide a quantitative analysis of these \sysname{}s to illustrate how they did not differ significantly from those authored by other researchers.

\subsubsection{Procedure}

The study was conducted through a popular video conferencing software.
After a brief introduction to the overall study, participants performed the task for each paper in order.
For each paper, participants were first provided with a short tutorial to the interface(s) that they would be using and, using a example paper and video, were allowed to use and test the interfaces for 5 minutes.
Then, participants proceeded to the assigned paper and were instructed to first fully read the paper's abstract.
After they read the abstract, participants were given 15 minutes to read the systems section of the paper and simultaneously write a summary that maximized the following criteria: 
\begin{itemize}
    \item \textit{Comprehensive}: how well the summary provides an overview of the entire section.
    \item \textit{Detailed}: how many specific details on the interactions and underlying models are included in the summary.    
    \item \textit{Coherent}: how well the summary flows or, in other words, how well the sentences connect logically. (This criteria was included to prevent summaries that simply listed details.)
\end{itemize}
To focus on capturing what they learned during the sessions, participants were informed that they could write a maximum of 14 sentences, were not allowed to copy-paste, and that the quality of their writing (e.g., spelling, grammar) would not be evaluated.
Once the given time passed, participants were asked to complete a survey about the task and, then, proceeded to the next task.
After all the tasks, we conducted a semi-structured interview about participants overall experience.

\subsubsection{Measures}

To evaluate the summaries, we developed a rubric for each paper where we listed all of the details contained in the system section of the paper, and we grouped these details according to the aspect of the system that they described (e.g., feature, pipeline component).
Then, for each summary, we annotated whether the summary presents each of these details and rated its coherency on a 7-point Likert scale.
To measure detail, we counted the number of details included in the summary, and, to measure comprehensiveness, we calculated the proportion of system aspects that were covered by the included details.
Two of the authors who did not observe the studies performed the annotations while being blind to the conditions that generated the summaries.
To verify reliability, the two authors first independently annotated the summaries for one paper, compared annotations and discussed to reach a consensus on the annotation process, and then independently annotated the paper again.
This resulted on a Cohen's kappa of 0.712 for annotating the details and Krippendorff's alpha of 0.744 for coherency ratings. 
As the agreement was substantial, each of the authors was assigned with one of the remaining papers, and they independently annotated the summaries for that paper.

\begin{figure}[!b]
    \centering
    \includegraphics[width=0.8\columnwidth]{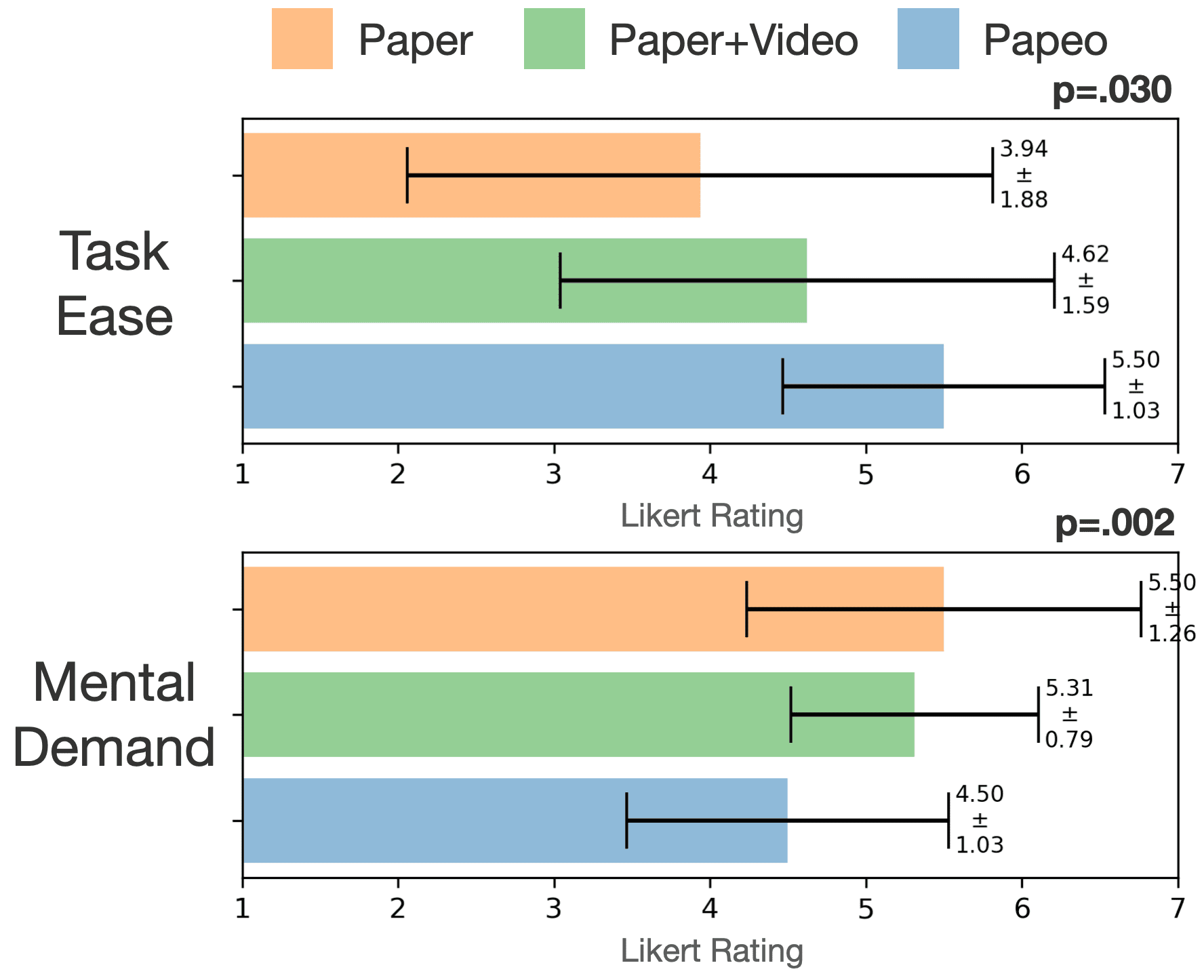}
    \caption{Perceived ease and mental demand were significantly affected by the condition used by participants. With \sysname{}s, task ease was perceived to be highest and mental demand the lowest.}
    \label{fig:ease-demand}
\end{figure}

Additionally, we collected participants ratings, on a 7-point Likert scale, to the following five questions from the survey: \textit{``I found it easy to write the summary''}, \textit{``I found it easy to orient myself (i.e., know where information is) in the paper/video''}, and \textit{``I found it easy to navigate to different parts of the paper/video''}.
The survey also contained five questions from the NASA-TLX questionnaire~\cite{hart1988development} to measure perceived workload---excluding the question on physical demand.
Finally, we analyzed interaction logs to measure how frequently participants (1) switched between the formats, (2) scrolled in the paper, and (3) scrubbed in the video. 
For switches, we counted every instance where the user interacted with one format after interacting with the other format, for scrolling and scrubbing, all consecutive actions within one second and in the same direction were counted as one action. 

\subsection{Results}

Our results revealed that \sysname{}s helped reduce participants' mental load during reading, facilitated and promoted navigation of both the paper and video, and led to more comprehensive summaries.
For the statistic analysis of each measure, we first conducted a Shapiro-Wilk test to determine if the data was parametric or non-parametric. 
Then, when comparing between all three conditions, we used a one-way, repeated measures ANOVA when parametric and a Friedman test when non-parametric
When comparing between the Paper+Video and \sysname{} conditions, we used a paired t-test when parametric and Wilcoxon signed-rank test when non-parametric.

\subsubsection{\textbf{Enhance Understanding and Decrease Mental Load}}

As seen in Figure~\ref{fig:ease-demand}, participants perceived the reading and summarizing task to be easiest with \sysname{}s.
The ANOVA analysis showed a significant effect of the condition on participants' perceived ease (Q=6.982, p=0.030) and a gradual increase between conditions, with the task perceived to be easiest in the \sysname{} condition.
This indicates that talk videos could facilitate the task for participants, but the support was not perceived as significant until they were integrated into the reading experience in \sysname{}s.
This is also reflected by responses to the NASA-TLX questionnaire as there was significant effect of the conditions on mental demand (Q=12.182, p=0.002) and demand was perceived to be lowest when participants used \sysname{}s.
Furthermore, although these results were not significant, perceived temporal demand, effort and frustration were lowest and perceived performance was highest with \sysname{}s (\autoref{tab:nasatlx_joint}). 

\begin{table}[!t]
\begin{tabular}{@{}lccccc@{}}
\toprule
\textbf{Condition} &
  \textbf{Mental} &
  \textbf{Temp.} &
  \textbf{Effort} &
  \textbf{Perf.} &
  \textbf{Frus.} \\ \midrule
Paper &
  \begin{tabular}[c]{@{}c@{}}5.50 \\ (1.27)\end{tabular} &
  \begin{tabular}[c]{@{}c@{}}5.13 \\ (1.63)\end{tabular} &
  \begin{tabular}[c]{@{}c@{}}5.25 \\ (1.48)\end{tabular} &
  \begin{tabular}[c]{@{}c@{}}4.44 \\ (1.46)\end{tabular} &
  \begin{tabular}[c]{@{}c@{}}4.00 \\ (1.79)\end{tabular} \\
\begin{tabular}[c]{@{}l@{}}{Paper +}\\ {Video}\end{tabular} &
  \begin{tabular}[c]{@{}c@{}}5.31\\ (0.79)\end{tabular} &
  \begin{tabular}[c]{@{}c@{}}5.19\\ (1.38)\end{tabular} &
  \begin{tabular}[c]{@{}c@{}}5.25\\ (0.93)\end{tabular} &
  \begin{tabular}[c]{@{}c@{}}4.63\\ (1.15)\end{tabular} &
  \begin{tabular}[c]{@{}c@{}}3.88\\ (1.15)\end{tabular} \\ 
Papeo &
  \begin{tabular}[c]{@{}c@{}}\textbf{4.50} \\ (1.03)\end{tabular} &
  \begin{tabular}[c]{@{}c@{}}\textbf{4.38}\\ (1.63)\end{tabular} &
  \begin{tabular}[c]{@{}c@{}}\textbf{4.63}\\ (1.31)\end{tabular} &
  \begin{tabular}[c]{@{}c@{}}\textbf{4.94}\\ (1.39)\end{tabular} &
  \begin{tabular}[c]{@{}c@{}}\textbf{3.63}\\ (1.59)\end{tabular} \\ \midrule
p-value &
  \textbf{0.002} &
  0.230 &
  0.249 &
  0.355 &
  0.715 \\ \bottomrule
\end{tabular}
\caption{Mean and standard deviation (in parentheses) of NASA-TLX questionnaire responses on mental demand, temporal demand, effort, performance, and frustration. (n=48, p-value based on Friedman tests.)}
\label{tab:nasatlx_joint}
\end{table}

\begin{figure*}[!b]
    \centering
    \includegraphics[width=0.80\textwidth]{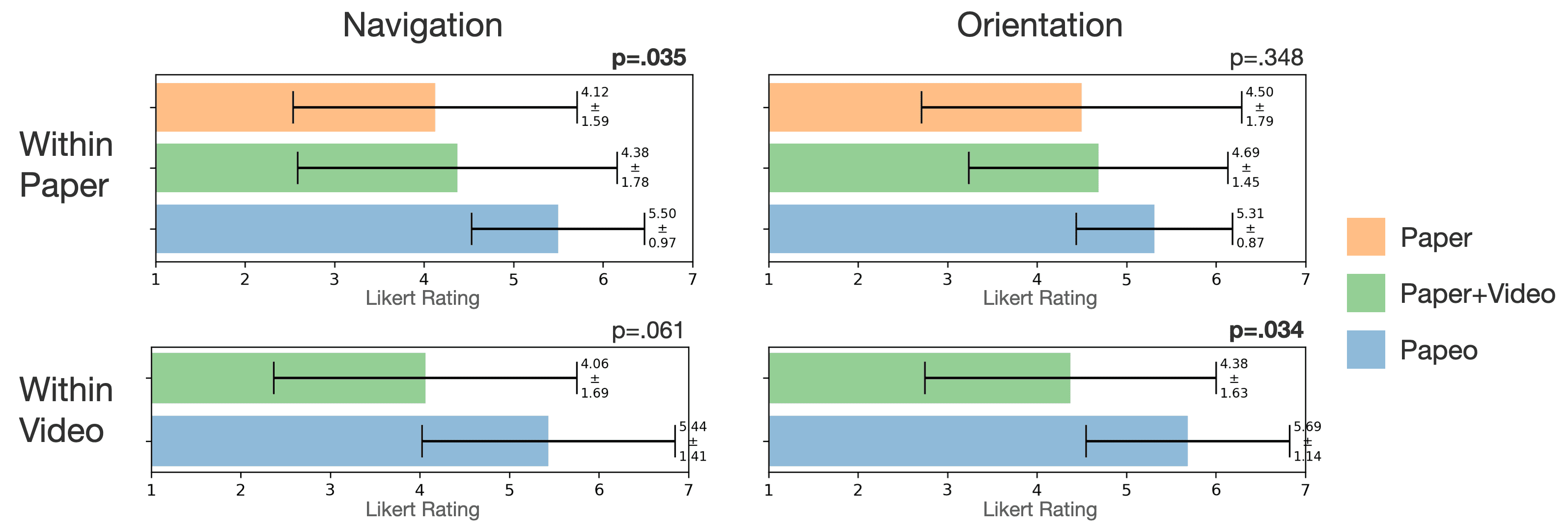}
    \caption{Results for perceived ease of navigation and orientation within the paper and the video. The condition had a significant effect on navigation within the paper and orientation within the video, with both perceived to be easiest with \sysname{}s.}
    \label{fig:navigation-results}
\end{figure*}

According to participants' comments, these results could be attributed to the various ways (i.e., summaries, modalities, alternative explanations) in which talk videos supported understanding and how \sysname{}s made these benefits available on demand.
For example, P14 mentioned how \sysname{}s summarized dense technical details but granted access to these details if needed: \textit{``The video is high-level summary. It was easier to understand and, if I need to understand technical details, I can look the highlighted section.''}
Additionally, P12 mentioned how \sysname{}s allowed them to combine and consume multiple modalities simultaneously: \textit{``Absolutely [preferred \sysname{}s] because I was visualizing and hearing the voice and reading the text. It was like three senses were active.''}
Finally, beyond helping them understand, P8 described how \sysname{}s allowed them to check their understanding by listening to alternative explanations: \textit{``English is not my first language so sometimes I will have a concern whether I understand the author’s intention correctly. But, with the video, usually they will discuss their research in more informal way.''}

\subsubsection{\textbf{One Format as a Guide for the Other}}

As \sysname{}s linked papers and videos, participants were able to use one format to guide their exploration of the other (Fig.~\ref{fig:navigation-results}).
Specifically, we observed that the condition had a significant effect on participants' perceived navigation ease within the paper (Q=6.704, p=0.035), where participants perceived it to be easiest with \sysname{}s and similar in the Paper and Paper+Video conditions.
According to participants, the links between the paper and video in \sysname{}s allowed them to navigate at a more fine-grained level than it was possible through the typical features of a paper.
P11 mentioned, \textit{``It breaks down the structure of the paper even more than the subsection headings. It also allows me to easily look for further details in the paper.''}
Additionally, P16 described how they were able to \textit{``move through the paper seamlessly''} by navigating according to the video notes through the autoplay feature.

In the opposite direction, participants perceived that it was significantly easier to orient themselves within the video in the \sysname{} condition when compared to the Paper+Video condition (W=27.000, p=0.034).
This signifies that it was easier for participants to know and remember where specific information was found within the talk video when using \sysname{}s.
P2 described that, with \sysname{}s, it was \textit{``clear which part of video [was] linked to''} to a specific passage of the paper, making it easier for them to find information they needed from the video.
Through the localized video notes, participants could immediately access video segments that they needed when they needed them---without searching for them through the video.
Thus, in \sysname{}s, the video supported navigation in the paper, and the paper supported orientation in the video.

\subsubsection{\textbf{Explore Easily, Engage More}}

Our analysis of the interaction logs (Fig.~\ref{fig:navigation-results}) revealed that participants engaged more with both formats when using \sysname{}s.
Participants switched between formats significantly more frequently in the \sysname{} condition compared to the Paper+Video condition (W=0.000, p<0.000). 
During the study, we observed that, in the Paper+Video condition, most participants watched the whole video first and then focused only on the paper during the remaining duration of the task.
However, in the \sysname{} condition, participants continuously switched back-and-forth between the formats.
Our analysis also revealed that the condition had a significant effect on how much participants scrolled in the paper (F=7.065, p=0.003) with participants scrolling to a similar degree in the \sysname{} and Paper conditions, and scrolling less in the Paper+Video condition.
Additionally, participants scrubbed in the video to a similar degree in both the \sysname{} and Paper+Video conditions (t=-1.810, p=0.090).
Considering how participants considered that it was easier to navigate in the paper and orient oneself in the video with \sysname{}s, these results suggest that \sysname{}s encouraged participants to engage with both formats, and to seek for and leverage their content during the task.

\begin{figure}[!b]
    \centering
    \includegraphics[width=0.70\columnwidth]{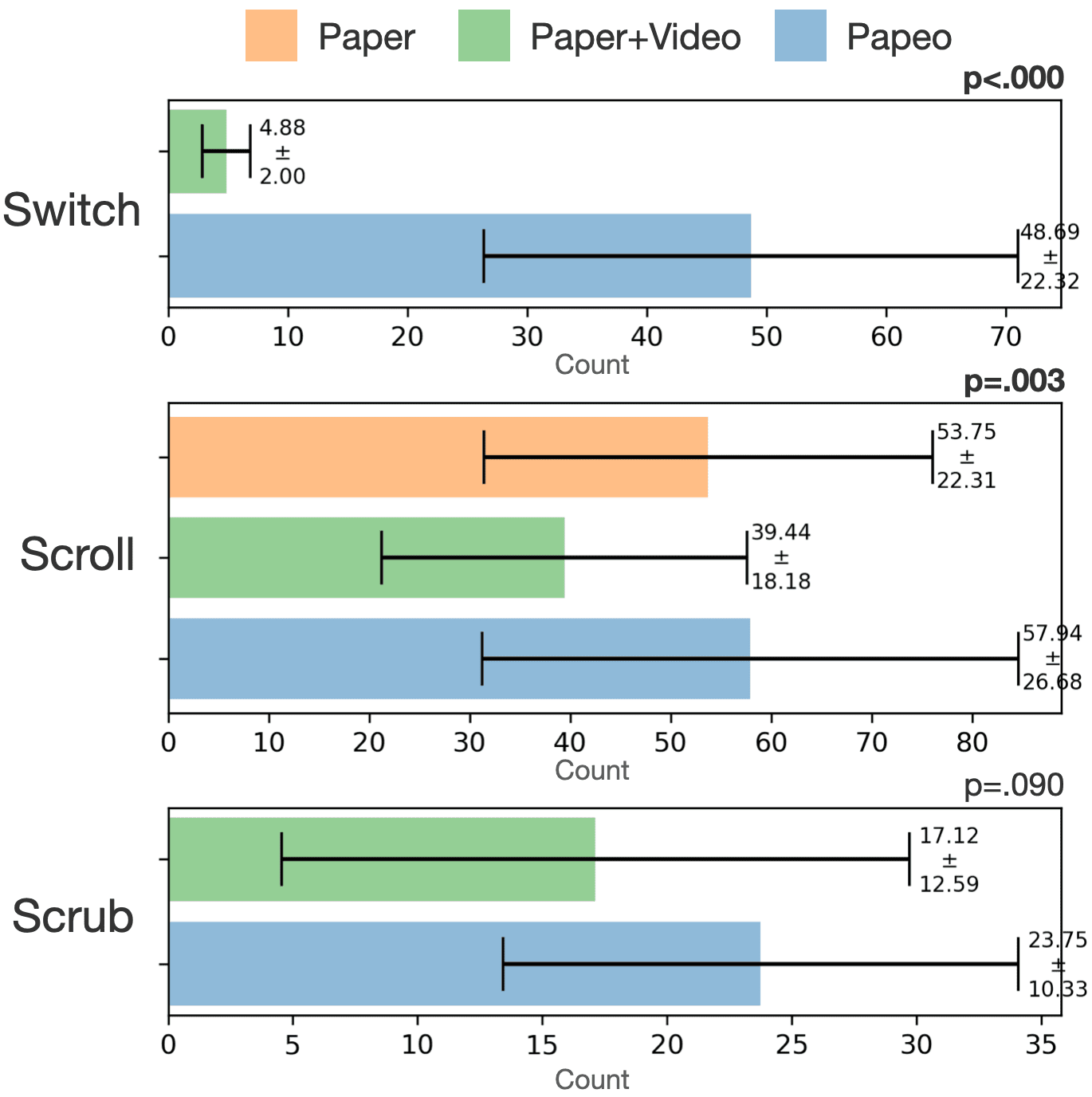}
    \caption{Analysis of the frequency of switching between formats, scrolling in the paper, and scrubbing in the video showed that the condition had a significant effect on switching and scrolling.}
    \label{fig:navigation}
\end{figure}

\subsubsection{\textbf{More Comprehensive Coverage}}

\begin{figure}[!b]
    \centering
    \includegraphics[width=0.85\columnwidth]{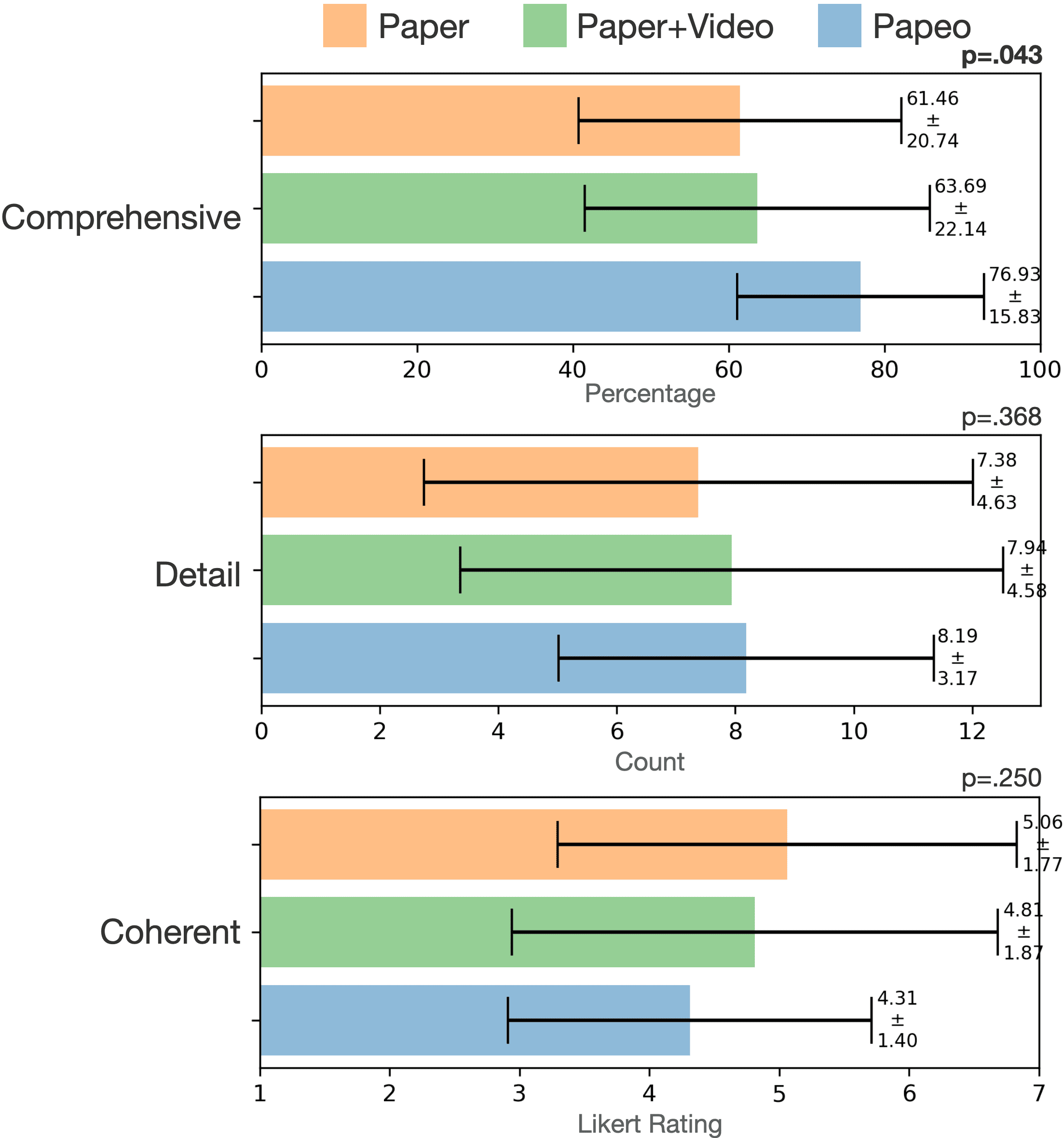}
    \caption{Results for the evaluation of participants' summaries according to comprehensiveness, detail, and coherency. The condition had a significant effect on comprehensiveness of the summaries, with summaries evaluated to be the most comprehensive in the \sysname{} condition.}
    \label{fig:summaries}
\end{figure}

The analysis of participants' summaries (Fig.~\ref{fig:summaries}) revealed that there was significant effect of the condition on the comprehensiveness of participants' summaries (F=3.497, p=0.043).
Summaries in the \sysname{} condition were rated to be the most comprehensive while those in the Paper and Paper+Video condition were relatively similar.
A plausible reason for this result is that, as \sysname{}s facilitated exploration of the content, participants were able to delve into details throughout the section and were thus able to include these in their summaries.
In terms of the other measures, there was no observed effect of the condition on the detail (Q=2.000, p=0.368) or coherency (Q=2.772, p=0.250) of participants' summaries.
This demonstrates that, despite participants interacting with both formats more with \sysname{}s and writing more comprehensive summaries, this was not at the expense of other qualities in participants' summaries.

\section{Field Deployment}

To further investigate how researchers would engage with \sysname{}s in the wild, we deployed this new format during CSCW 2022.
During the duration of the conference, we promoted our interface through social media channels and a daily newsletter sent to conference attendees.
Through a portal website, conference attendees could access our reading interface and consume \sysname{}s for specific papers that were being presented during the conference.
To pre-populate this set of \sysname{}s, we contacted several authors that were presenting in the conference and asked if they would like to use our authoring interface to create \sysname{}s to promote their papers.
Through this, we collected a set of 12 \sysname{}s (or around six hours of volunteered authoring).
The portal website also provided tutorials for using and creating \sysname{}s and described what data is collected by the interfaces.

During the two weeks of the conference, our reading interface was visited by 288 unique users and, on average, each user visited a total of 1.20 different \sysname{}s (min=1, max=5).
To analyze the interaction logs, we identified user sessions (i.e., sequence of actions between entering and leaving the interface) and removed anomalous sessions (e.g., user left the interface immediately after entering, or user entered the interface but only interacted with it hours later). We observed that readers were actively engaged with \sysname{}s.
The average number of actions per session (e.g., scroll, play video, scrub) was 32.02 (min=2, max=255) and the average session length was 5.74 minutes (min=1.02, max=40.21).
In addition to these statistics, various researchers expressed positive comments about \sysname{}s on social media.
One researcher expressed how \sysname{}s were \textit{``easily scannable and digestible''}, which reflected our study findings, and another researcher noted how the format can \textit{``humanize''} papers by letting the reader \textit{``hear the author's voice saying words that are often part of the fabric of the paper.''}
Beyond these benefits, a researcher noted how \sysname{}s can \textit{``do more than just replicate the print experience''} and \textit{``help so all the effort we put into presentation videos doesn't get completely buried after a conference''}.
In sum, through a field deployment, researchers found value in \sysname{} for real-world use cases, and wider adoption may require further lowering the cost of authoring \sysname{}s. 

\section{Discussion}

In this paper, we propose \sysname{}s, a novel reading experience that augments research papers with localized segments from talk videos to support skimming, navigation, and comprehension.
While designers and researchers have taken various steps to reach the vision of \textit{dynamic reading}, as discussed by Victor~\cite{victor2011explorable, victor2014humane}, the experiences they proposed required a prohibitive amount of effort to realize (e.g., authoring animations or demos~\cite{grossman2015your,masson2020chameleon}).
In fact, \textit{Distill}, a peer-reviewed journal that published interactive articles, cited authoring effort as a reason for their discontinuation~\cite{team2021distill}.
In our work, we instead recognize that researchers have already dedicated significant effort in authoring talk videos that may already possess features that can enhance academic papers, like progressive animations and demo walkthroughs.
Our \sysname{} experience leverages these talk videos to, with relatively minimal additional effort, augment the experience of reading academic papers---taking a step towards the vision of \textit{dynamic reading}.

To extend on this vision, we identify various directions for enhancing and expanding on \sysname{}s: automating the creation of \sysname{}s to expand their availability, extending to other types of videos or content (i.e., blog posts), and leveraging paper-video links to generate talk videos from papers.
In this section, we elaborate on the potential of \sysname{}s and on these directions for future work.

\subsection{Papeos Everywhere}

Through our user study, we identified that \sysname{}s can support understanding and navigation of papers---lowering various barriers of research consumption.
Although they can aid early-stage researchers to access a larger body of knowledge, the coverage of papers that are supported with \sysname{}s is limited by the paper authors' willingness to create \sysname{}s.
In our work, we focused on providing authors control over how their \sysname{}s are created due to their concerns regarding automation errors.
While this decision respects their preferences as authors, researchers as readers may desire a fully automatic approach as, despite possible errors, this enables them to leverage talk videos in more papers---a conflicting sentiment shared by various participants in the formative study. 
To increase the coverage of \sysname{}s, future work could further develop the AI-based pipeline used for suggestions in the authoring interface.
Specifically, the talk video segmentation algorithm can be enhanced by combining both visual and textual features.
Additionally, while our work used general-purpose, state-of-the-art text embedding models, a small-scale dataset of paper-video links could be collected to fine-tune a sentence transformer~\cite{reimers2019sentence} for this specific setting. 
However, as an improved pipeline may still present errors, the reading interface should be enhanced to provide users with error-recovery mechanisms---e.g., present multiple passages that could link to a video segment and allow the user to override erroneous links. 
With these improvements, future work can widen the availability of \sysname{}s and lower the floor for early-stage researchers.

\subsection{Beyond Talks and Videos}

While our work focused on augmenting papers with talk videos, researchers employ an assortment of varying formats to communicate their research, and these could also be adopted to augment papers.
For videos, there are various formats that exist aside from recordings of conference talks: video figures, demo videos, recordings of invited talks or thesis defenses, and, more recently, paper ``explainers'' on platforms like YouTube.\footnote{Example channels include \textit{Two Minute Papers} and \textit{AI Coffee Break}.}
These video formats may differ from talk videos and can therefore provide different benefits when employed in \sysname{}s. 
For example, demo videos can present systems and their features in more detail, invited talks or thesis defenses can contextualize a paper within a extended thread of work, and ``explainer'' videos can simplify the content further as their target audience can include non-researchers. 
Instead of depending on existing videos, authors could also create custom video clips to augment their papers with \sysname{}s in different forms.
For example, while talk videos are constrained in length and were thus useful to summarize and skim the paper, custom video clips would not be constrained and may allow authors to augment their papers with extensive, additional commentary or comprehensive walkthroughs of interfaces.

Aside from the visual aspect of videos, study participants and users from the deployment noted the significance of incorporating audio into the papers: enabling consumption with various modalities and ``humanizing'' papers.
Future iterations of \sysname{} could support authors in creating additional audio-based notes to weave their voices into their papers.
As an additional benefit, these audio notes could supplement screen readers and help increase the accessibility of papers by providing authors with a lightweight mechanism for creating alternative descriptions.
Beyond videos, researchers frequently promote their research through other channels such as blog posts and social media (e.g., Twitter threads), and \sysname{}s could integrate content from these formats as text-based notes.
As research is increasingly distributed through a greater number of formats, \sysname{}s can serve as a first step to connect these forms into one cohesive experience. 

\subsection{Generating Videos for Papers}

As talk videos only cover a subset of the paper, \sysname{}s can surface the important passages of the paper but, due to the same reason, they cannot provide video notes for the other passages.
In our user study, several participants expressed how they could struggle to understand a passage, but were disappointed to not find any video notes to assist them. 
To remedy this limitation, future work could extend on existing work on document-to-video generation~\cite{chi2021automatic} to automatically generate video segments from paper passages.
Specifically, with passages as input, a pipeline could generate summaries for the video's transcript~\cite{lev2019talksumm} and slides for the frames~\cite{fu2022doc2ppt, wang2023slide4n}. 
Then, the pipeline could produce video segments by combining these and incorporating audio with text-to-speech models---or even add an artificial talking head~\cite{chi2022synthesis}.
To train and tune the AI models involved in this generative pipeline, future work could use our authoring interface to collect a larger dataset of paper passage and video segment pairs.
By presenting these generated video segments when requested by the reader, future \sysname{}s can more comprehensively support the paper reading experience.

\subsection{Limitations}

Our studies revealed various benefits of \sysname{}s that we believe can be generalize beyond the set of papers we have tested. At the same time, we acknowledge several factors could effect the usefulness of \sysname{}s: 
\begin{itemize}
    \item Type of work: Formative study participants noted that videos were more useful for work involving interactive and/or dynamic artifacts (e.g., HCI systems).
    \item Paper sections covered: User study participants expressed how \sysname{}s were especially helpful for summarizing information dense sections.
    \item Visuals: Formative and user study participants noted that supplemental visuals in videos, especially those animated or presented gradually, were effective illustrating information in the paper.
    \item Communication style: Formative and user study participants appreciated videos that communicated paper content in a different style (e.g., informal language).
\end{itemize}
Future work should investigate the effectiveness of our approach according to these factors.
Additionally, to fit the user study within 90 minutes, our user study focused on HCI papers with system contributions and only investigated the benefits of \sysname{}s when reading one section in the paper.
However, we argue that our various studies together demonstrated benefits of our approach that can generalize across papers, types of work, and domains: highlights, summaries, and audio narrations. 
For example, even for a qualitative paper, our approach can highlight important paper fragments (e.g., author selected themes and quotes), and provide the authors' audio narrations and summaries.
Future work can conduct additional studies to investigate the significance of these benefits with papers of diverse domains and contributions.

\section{Conclusion}

This paper presents \sysname{}s, a novel reading experience that integrates segments from talk videos as localized margin notes in academic papers.
To facilitate the creation of \sysname{}s, we introduce an authoring interface that aids paper authors in linking video segment and paper passages through algorithmic and AI-based suggestions. 
Through a within-subjects user study (n=16), we found that \sysname{}s could enhance understanding of papers by providing summaries of complex passages and allowing readers to consume information in multiple modalities.
With \sysname{}s, participants leveraged each format (i.e., paper and video) to guide their navigation in the other format, which in turn facilitated navigation in both formats and encouraged more comprehensive reading of the paper.
These findings and responses from researchers in a field deployment suggest the potential for leveraging existing, alternative forms of research communication to augment research papers and enable more dynamic reading experiences.
\begin{acks}

The authors would like to thank Doug Downey, Shannon Zejiang Shen, Evie Cheng, and Juho Kim for their insightful discussions and feedback. We also thank the anonymous reviewers for their constructive feedback. Finally, we would like to thank the various researchers that participated in our various studies and the deployment of our system.

\end{acks}

\bibliographystyle{ACM-Reference-Format}
\bibliography{references}

\clearpage
\appendix

\section{Quantitative Analysis of \sysname{}s}
\label{appendix:papeo-analysis}

We quantitatively analyzed various characteristics of all the \sysname{}s authored throughout our work (N=23). 
These include 8 \sysname{}s from the test set, 6 from the preliminary user evaluation of the authoring tool, and 9 additional ones from the deployment study. Although there was a total of 12 \sysname{}s in the deployment, 3 of them were authored during the preliminary evaluation of the authoring tool.
Table~\ref{tab:papeo-analysis-results} shows that the chosen \sysname{}s fall within one standard deviation for all of the characteristics, which suggests that they did not deviate significantly from those authored by other researchers.

\begin{table*}[!b]
\begin{tabular}{@{}lccccc@{}}
\toprule
\textbf{Characteristics}                                                              & \textbf{Mean} & \textbf{SD} & \textbf{\cite{kim2022stylette}} & \textbf{\cite{chang2021tabsdo}} & \textbf{\cite{huh2022cocomix}}  \\ 
\midrule
Number of Linked Paper Fragments and Video Segments                  
& 20.6  & 7.3         & 20            & 15            & 24             \\
Average Number of Paper Fragments per Link 
& 3.2  & 1.6          & 3.9           & 2.8           & 1.8            \\
Average Length of Linked Video Segment            
& 24.3  & 8.7         & 24.0          & 29.2          & 19.0           \\ 
Total Number of Synchronized Highlights          
& 2.8  & 4.2          & 7             & 7             & 3              \\ 
\bottomrule
\end{tabular}
\caption{Analysis of various characteristics of the \sysname{}s collected during this work. The analysis shows the overall statistics (i.e., mean and standard deviation) for all the \sysname{}s, and the statistics for the three that were selected for the user study.}
\label{tab:papeo-analysis-results}
\end{table*}

\end{document}